\documentclass[twocolumn]{aastex631}

\graphicspath{{./}{figures/}}
\usepackage{subfigure}

\usepackage{appendix}

\begin{document}

\title{High-resolution Elemental Abundance Measurements of Cool JWST Planet Hosts Using AutoSpecFit: An Application to the  Sub-Neptune K2-18b's Host M dwarf}

\author[0000-0001-5541-6087]{Neda Hejazi}
\affil{Department of Physics and Astronomy, University of Kansas, Lawrence,  KS 66045, USA}
\affil{Department of Physics and Astronomy, Georgia State University, Atlanta, GA 30303, USA}
\email{nhejazi@ku.edu}

\author[0000-0002-1835-1891]{Ian J. M. Crossfield}
\affil{Department of Physics and Astronomy, University of Kansas, Lawrence,  KS 66045, USA}

\author[0000-0002-7883-5425]{Diogo Souto}
\affiliation{Departamento de F\'isica, Universidade Federal de Sergipe, Av. Marcelo Deda Chagas, S/N Cep 49.107-230, S\~ao Crist\'ov\~ao, SE, Brazil}

\author[0000-0002-2072-6541]{Jonathan Brande}
\affil{Department of Physics and Astronomy, University of Kansas, Lawrence,  KS 66045, USA}

\author[0000-0001-5344-8069]{Thomas Nordlander}
\affil{Research School of Astronomy \& Astrophysics, Australian National University, Canberra, ACT 2611, Australia}
\affil{The ARC Centre of Excellence for All Sky Astrophysics in 3 Dimensions, Canberra, ACT 2611, Australia}

\author[0000-0001-8907-4775]{Emilio Marfil}
\affil{Departamento de F\'{i}sica de la Tierra y Astrof\'{i}sica and IPARCOS-UCM (Unidad de F\'{i}sica de Part\'{i}culas y del Cosmos de la UCM),
Facultad de Ciencias F\'{i}sicas, Universidad Complutense de Madrid, 28040 Madrid, Spain}
\affil{Hamburger Sternwarte, Universit\"{a}t Hamburg, Gojenbergsweg 112, 21029 Hamburg, Germany}

\author[0000-0001-6476-0576]{Katia Cunha}
\affil{Department of Astronomy and Steward Observatory, University of Arizona, Tucson, AZ 85721, USA}

\author[0000-0002-1221-5346]{David R. Coria}
\affil{Department of Physics and Astronomy, University of Kansas, Lawrence,  KS 66045, USA}

\author[0000-0002-0475-3662]{Zachary G. Maas}
\affil{Department of Astronomy, Indiana University, Bloomington, IN 47405, USA}

\author[0000-0001-7047-8681]{Alex S. Polanski}
\affil{Department of Physics and Astronomy, University of Kansas, Lawrence,  KS 66045, USA}

%\author[0000-0002-7615-4028]{Yakiv Pavlenko}
%\affil{Instituto de Astrof\'isica de Canarias, E-38205 La Laguna, %Tenerife, Spain}

\author[0000-0003-0595-5132]{Natalie R. Hinkel}
\affil{Department of Physics and Astronomy, Louisiana State University, Baton Rouge, LA 70803, USA}

\author[0009-0004-9078-5987]{Joseph E. Hand}
\affil{Department of Physics and Astronomy, University of Kansas, Lawrence,  KS 66045, USA}

\begin{abstract}
We present an in-depth, high-resolution spectroscopic analysis of the M dwarf K2-18 that hosts a sub-Neptune exoplanet in its habitable  zone. We show our technique to accurately normalize the observed spectrum, which is crucial for a proper spectral fitting. We also introduce a new automatic, line-by-line model-fitting code, AutoSpecFit, that performs an iterative {\ensuremath{{\chi}^2}} minimization process to measure individual elemental abundances of cool dwarfs. We apply this code to the  star K2-18, and measure the abundance of 10 elements - C, O, Na, Mg, Al, K, Ca, Sc, Ti, and Fe. We find  these abundances  moderately supersolar, except for Fe with a slightly subsolar abundance. The accuracy of the inferred abundances is limited by the systematic errors due to uncertain stellar parameters.  We also derive  the abundance ratios associated with several planet-building elements such as  Al/Mg, Ca/Mg, Fe/Mg, and (a solar-like) C/O=0.568 $\pm$ 0.026, which can be used to constrain the chemical composition and the formation location of the exoplanet. On the other hand, the planet K2-18 b has attracted considerable interest, given the JWST measurements of its atmospheric composition. Early JWST studies reveal an unusual chemistry for the  atmosphere of this planet, which is unlikely to be driven by formation in a disk of unusual composition. The comparison between the chemical abundances of K2-18 b from future JWST analyses and those of the host star can  provide fundamental insights into the formation of this planetary system.

\end{abstract}


\section{Introduction} \label{sec:intro}
Cool dwarfs ({\ensuremath{M{\lesssim}0.75M_{\sun}}}) are optimal and primary targets for transit and radial velocity surveys of planets beyond our Solar system, since their lower mass, radius, and luminosity make planetary signatures easier to detect compared to those exoplanets orbiting more massive dwarfs. M dwarfs ({\ensuremath{M{\lesssim}0.6M_{\sun}}}) are particularly the most abundant stars in the Galaxy ({\ensuremath{70\%}} by number, \citealt{Reid_Gizis1997, Henry2006}), and there is likely at least one planet orbiting around these stars \citep[e.g.][]{Dressing_Charbonneau2013, Dressing_Charbonneau2015, Tuomi2014, Hardegree-Ullman2019}. M dwarfs therefore dominate the general occurrence rates of planets around main sequence stars. The presence and properties of planets are believed to be linked to the chemical composition of their host stars \citep[e.g.][]{Santos2004, Fischer_Valenti2005, Beauge_Nesvorny_2013}. Accordingly, M dwarfs provide ideal sites to probe the formation mechanisms of planetary systems.

Planets are formed in a protoplanetary disk around a new star, which are all embedded in a larger molecular cloud. As a result, there is a mutual interaction between planets and their host stars,  which can alter the properties of the two components over their lifetimes. In particular, the accretion of material from protoplanetary disk into the star as well as post-formation events, such as planet engulfment, may imprint planetary signatures in stellar chemical abundances \citep[e.g.][]{Pinsonneault2001, Oh2018,  Nagar2020, Spina2021}. The detailed abundance measurements of host stars are, therefore, of vital importance to characterizing planetary systems and can provide fundamental insights into planetary formation, evolution, and composition. 

Although significant progress has been made in understanding star-planet chemical connections,  most studies have been focused on more massive FGK-type dwarfs rather than M dwarfs. The spectra of cool M dwarfs are dominated by millions of molecular lines in both the optical (e.g., TiO and CaH) and near-infrared (NIR, e.g., H$_{2}$O) regions, which are blended with each other and many atomic lines. This causes a significant flux depression and, in turn, makes identifying the continuum level in many wavelength areas challenging. Combined with the substantial line crowding, established methodologies for FGK-type dwarfs and giant stars, such as equivalent width measurements, are therefore inappropriate for M dwarfs. As a result, most spectroscopic studies of M dwarfs rely on the spectral synthesis and model fitting \citep[e.g.][]{Rajpurohit2014, Lindgren2016, Souto2022}.

Recently, high-resolution NIR spectroscopy has opened the way for detailed elemental abundance measurements of M dwarfs with a reasonable accuracy ($\lesssim$ 0.2 dex). Modern spectrographs, along with methods to correct spectra for telluric contamination, have made it possible to detect and analyze various atomic and molecular lines and scrutinize the effect of physical parameters on their shape. Parallel advances in modeling the atmospheres of low-mass M dwarfs and calculating atomic and molecular line lists are of great importance in measuring the parameters and chemical abundances of these stars. Various previous studies have attempted to model M-dwarf atmospheres assuming one-dimensional radiative-convective equilibrium \citep[e.g.][]{Allard1995, Tsuji1996, Gustafsson2008, Kurucz2011, Hesser2013}. 
However, the synthetic spectra associated with the same set of physical parameters and elemental abundances using different model atmospheres and spectral synthesis methods show discrepancies over many wavelength ranges. These are likely due to differences in model assumptions and  opacity calculations as well as atomic and molecular line lists incorporated in synthesizing spectra (\citealt{Iyer2023}, specifically see Figure 1).  All these complications motivate more profound and detailed studies to understand any missing source of line and continuum opacity and better characterize the atmosphere of M dwarfs. Nevertheless, significant progress has been made in determining the physical parameters of M dwarfs using high-resolution NIR spectroscopy with synthetic model fitting  \citep[e.g.][]{Lindgren2016, Lindgren2017, Rajpurohit2018, Passegger2018, Passegger2019, LopezValdivia2019, Souto2017, Souto2020, Souto2021, Souto2022, Marfil2021, Wanderley2023} using different methods and various combinations of model atmospheres and line data. Although these studies have shown agreement between their own observed and best-fit model spectra, the consistency in parameter values among different analyses is still under debate \citep[e.g.][]{Olander2021, Passegger2022}.

In contrast to the numerous efforts aimed at the determination of M-dwarf physical parameters, measuring the individual elemental abundances of these cool dwarfs using line-by-line model fitting, particularly in high-resolution NIR spectra, is still in the early stage \citep[e.g.][]{Souto2017, Abia2020, Shan2021, Souto2022}. The accuracy of inferred elemental abundances from such methods highly depends on model atmospheres and atomic and molecular line lists used in spectral synthesis, as well as the continuum/pseudo-continuum normalization of observed spectra. \citet{Souto2017, Souto2022} derived the abundances of 13-14 elements for a few M-dwarf samples by synthesizing spectra using the widely-used radiative transfer code Turbospectrum  \citep{AlvarezPlez1998, Plez2012} along with one-dimensional (1D) plane-parallel MARCS model atmospheres \citep{Gustafsson2008}, and then performing a {\ensuremath{{\chi}^2}} minimization approach for each single selected spectral line. In our previous work (\citealt{Hejazi2023}, hereafter Paper I), we further extended this method by carrying out an iterative {\ensuremath{{\chi}^2}} minimization process, where after each iteration, a new grid of synthetic spectra for each element was generated based on the new inferred abundances, which were then used in the next iteration. This procedure was repeated until the abundance of all elements converged to their final values. In Paper I, the transition from one iteration to the next was implemented manually, but we have  developed a model fitting code, AutoSpecFit, that automatically allows Turbospectrum to produce the synthetic spectra required for each iteration ``on the fly'' without interrupting the run. In this paper, we apply this automatic code to the planet-hosting M dwarf  K2-18 to obtain its elemental abundances. The sub-Neptune K2-18b has been targeted by several James Webb Space Telescope (JWST) observing programs, and the comparison between the composition of this planet and its host star can shed light on its formation history.

This paper is outlined as follows. In Section \ref{sec:exok2_18b}, we describe the properties of the exoplanet K2-18 b that has been observed by both the Hubble Space Telescope (HST) and JWST. We summarize  the  spectroscopic observations of the host star K2-18, the data reduction method, and the pre-processing needed to prepare the spectra for the analysis in Section \ref{sec:obs}. The spectral synthesis and the line lists used in this study are described in Section \ref{sec:spec_synthesis}. The process of line selection and continuum/pseudocontinuum normalization are presented in Section \ref{sec:line_normalize}. Physical parameters of the target K2-18 determined from other independent methods, except for microturbulent velocity that is obtained from this spectroscopic analysis,  are shown in Section \ref{sec:all-k2-18-par}. All steps of AutoSpecFit for measuring elemental abundances are detailed in Section \ref{sec:autospecfit}. In Section \ref{sec:application}, we utilize our abundance technique to derive  the abundances of  10  elements as well as the abundance ratios associated with several planet-building elements for K2-18. The error analysis of the inferred abundances and abundance ratios is also demonstrated in this section. The summary and conclusion of this study, particularly in the context of the star-planet connection, are presented in Section \ref{sec:summary}.

\section{Exoplanet K2-18 \MakeLowercase{b}}  \label{sec:exok2_18b} 

The K2-18 system is host to two planets, one of which (K2-18 b) is a transiting super-Earth/sub-Neptune (2.61 $\pm$ 0.09 $R_\oplus$, $9.51^{+1.57}_{-1.89}$ $M_\oplus$\footnote{The earlier study of \cite{Cloutier2017} inferred a planet mass of 8.63 $\pm$ 1.35 $M_\oplus$, which is consistent with the planet mass from \cite{Radica2022} within the uncertainties.}; \citealt{Benneke2019, Radica2022}) in the star's habitable zone \citep{Montet2015, Crossfield2016}, and the other (K2-18 c) is a non-transiting planet of similar mass \citep[$6.92^{+0.96}_{-0.99}$ $M_\oplus\sin i$;][]{Radica2022}. Given K2-18 b's amenability to transit spectroscopy and its temperate instellation, it has been a high-priority target for observational and theoretical characterization. 

Initial HST/WFC3 transmission spectroscopy revealed a clear atmosphere for the planet, as well as the presence of water vapor \citep{Benneke2019, Tsiaras2019}. The prospect of water vapor on a habitable zone world spurred a flurry of further modeling to explain the observed data and more thoroughly model the planet's upper atmosphere and deeper interior. \cite{Madhusudhan2020} modeled the interior structure of the planet and how varying interior compositions would affect the planet's observed spectrum, finding that, while their modeling efforts are consistent with rocky planets, classical gas dwarfs, and water-dominated ocean worlds, K2-18 b is likely to have a small ($\lesssim6\%$) H/He fraction, and the planet could still support habitable conditions. \cite{Bezard2022} noted that methane has strong absorption features that overlap with water vapor in the HST/WFC3 near-IR bandpass and found that, after reanalyzing the data, methane is a much more likely absorber given self-consistent radiative-equilibrium models of K2-18 b's atmosphere. 

This predicted methane signal was confirmed with JWST observations of the planet, clearly detecting methane and carbon dioxide (and not detecting water vapor) at wavelengths without contaminating features from other absorbers \citep{Madhusudhan2023}. Again, many more theoretical investigations followed this reversal of the previous observational results, focusing on the potential for K2-18 b to be a ``Hycean'' (water ocean under a hydrogen atmosphere) planet compared to a more typical Neptune-like gas-dwarf. By modeling the convective processes on K2-18 b, \cite{Leconte2024} predict that the planet may not be Hycean, as its clear atmosphere would allow too much incident radiation to maintain a liquid water ocean, while \cite{Shorttle2024} show that a magma ocean interior could also reproduce the current observed JWST spectrum. 

Finally, \cite{Wogan2024} model the planet and favor a typical Neptune-like composition over Hycean compositions as Hycean planets may not be able to produce sufficient methane through photochemical processes to match the observed methane abundance in the JWST data. Other similar exoplanets have also been observed in the same mass/radius/temperature range as K2-18 b, such as TOI-270 d, another habitable-zone sub-Neptune with methane and carbon dioxide, but also water vapor \citep{Benneke2024, Holmberg2024}. The persistent uncertainties around K2-18 b's true nature and the infancy of panchromatic, high-precision studies of these temperate worlds both motivate deeper studies of the system itself, especially this work.

\section{Spectroscopic Observations and Pre-processing}  \label{sec:obs}
We observed K2-18 with the IGRINS high-resolution (R$\sim$45,000) spectrograph \citep{Yuk2010,Park2014} at the Gemini-South Observatory as part of program GS-2023A-Q-203 (PI: Ian Crossfield).  The star was observed on 2023-01-20 with a single ABBA nod sequence; each frame had an exposure time of 245~s.  For telluric corrections, the facility observers selected the nearby A0V star HIP~61628 and observed a single ABBA sequence with 50~s integration times.  The data were processed in the same way as described in Paper I.  In brief, the raw 2D echelleograms were processed and reduced by the standard IGRINS Pipeline Package \citep{Lee2017}, with the order-by-order 1D spectra provided through the Raw \& Reduced IGRINS Spectral Archive \citep{Sawczynec2022}.  We then further processed the spectra by running the spectra of K2-18 and its A0V calibrator through the \texttt{SpeXTool} pipeline's \texttt{xtellcor\_general} routine \citep{Cushing2004} to account for any small wavelength offset between the spectra of K2-18 and the A0V star, and then through \texttt{xmerge\_orders} to combine the individual echelle orders into a single, 1D spectrum.  The final spectrum spans a wavelength range of 1.45-2.45\,{\micron} covering both H- and K- bands, with a median S/N of  270 per pixel, which is higher than the minimum median S/N ($\sim$200)  required for detailed abundance measurements of cool dwarfs at the resolution provided by IGRINS spectra (or even at the lower resolution of APOGEE spectra, i.e., $\sim$22,500).

In order to flatten the observed spectrum, we divide the spectrum into smaller parts, typically ranging between 50 and 150 {\AA}, and fit a low-order polynomial to the data points of each part. We then exclude those wavelengths whose fluxes are less than the respective values of the polynomial. We further fit a new low-order polynomial to the remaining data points and again exclude those wavelengths with fluxes less than the relevant polynomial values. This procedure is repeated until a final polynomial, that passes only through the spectral peaks and does not cross any absorption line, is reached. Lastly, we divide the spectrum of each part to the corresponding final polynomial to obtain the flattened spectrum normalized to unity, and then combine all the flattened parts back together. It should be noted that the resulting flattened spectrum does not present a continuum-normalized spectrum as the continuum level of  M-dwarf spectra cannot be identified in many spectral regions.

\section{Spectral Synthesis and Line Data}  \label{sec:spec_synthesis}
We generate the synthetic, continuum-normalized spectra\footnote{The synthetic continuum-normalized spectra are calculated by dividing the absolute flux line spectrum by the absolute flux continuum spectrum. The continuum is calculated in the same way as the line spectrum, but instead of line opacities, only continuous opacities are used. This approach is a standard practice in high-resolution spectroscopic analyses, as the continuum generally exhibits smooth variations on scales longer than the width of a spectral order. The continuum is calculated on a more coarse wavelength scale than the line spectrum, and then interpolated onto the exact same wavelengths.} (hereafter, ``synthetic models/spectra'' or ``model spectra'', for simplicity) required for our analysis by employing Turbospectrum (version v15.1) assuming local thermodynamic equilibrium (LTE)\footnote{The non-LTE version of Turbospectrum \citep{Gerber2023} has also been publicly available.}, which can consistently handle very large line lists including millions of spectral lines related to all atoms and tens of molecules. We use 1D hydrostatic MARCS model atmospheres that were computed in LTE and solves the radiative transfer equation in plane-parallel geometry for dwarf stars. The MARCS model grid is based on the solar abundances from \citet{Grevesse2007}, but the abundances of $\alpha$-elements are enhanced for models with subsolar metallicities ([M/H]$<$0) following the typical trends of [$\alpha$/Fe] as a function of metallicity for stars in the Solar neighborhood. To synthesize model spectra, we also use a set of atomic and molecular line lists that are described in Paper I, but with some improvements, as shown below.

To examine our selected atomic line list (and also to choose the best spectral lines and perform the pseudo-continuum normalization process, Section \ref{sec:line_normalize}), we need to compare our observed spectrum to an initial guess of best-fit model. To this end,  We use the interpolation routine  developed by Thomas Masseron\footnote{https://marcs.astro.uu.se/software.php} to interpolate the MARCS model associated with the star's physical parameters (see Section \ref{sec:k2-18-par}). Using the interpolated model,  we then produce the synthetic spectrum  with the star's parameters, assuming microturbulence $\xi$=1.0 km/s,  and the absolute abundances equal to the solar values plus the overall metallicity, i.e., {{A(X)=A(X)$_{\sun}$+[M/H]}}, or equivalently, the relative abundances\footnote{In this paper, we use several abundance notations that are defined as follows: \\
``absolute abundance'' $\rm {A(X)=log({N_{X}}/{N_{H}})_{star}+12}$, \\
``abundance'' $\rm{[X/H]=log({N_{X}}/{N_{H}})_{star}-log({N_{X}}/{N_{H}})_{\sun}}$, \\
or  $\rm{[X/H]=A(X)_{star}-A(X)_{\sun}}$, \\
``relative abundance'' [X/Fe]=[X/H]$-$[M/H], \\
``abundance ratio'' $\rm{X/Y=10^{(A(X)-A(Y))}=N_{X}/N_{Y}}$, \\
where X and Y indicate the elements X and Y, $\rm {N_{X}}$ means the number density of element X, and [M/H] shows the overall metallicity.} equal to the solar values, i.e., [X/Fe]=0, where X denotes the element X. These solar relative abundances are the default values when using Turbospectrum without any abundance customization. Although we first assume a microturbulence value of $\xi$=1.0 km/s based on  previous studies of M dwarfs \citep[e.g.][]{Souto2017}, we later find this value as the best-fit  parameter for K2-18 (see Section \ref{sec:vmic}). This synthesized spectrum represents a first-order approximation of the star's best-fit model (hereafter ``(Model){$_{\textrm{approx}}$'').

For Radial velocity correction, we compare this model with the observed spectrum that is Doppler shifted to obtain the star's radial velocity. We first visually examine different radial velocities (RVs) with  a large step of 10.0 km/s over several spectral regions, and after finding a rough estimate of RV, we  determine the best-fit RV value by fine tuning using  small steps between 0.5 and 1.0 km/s. However, smaller RV adjustments which can be as small as $\pm$ 0.1 km/s) may still be needed for some spectral lines before synthetic model fitting. This slight radial velocity offset may be due to the uncertainty of the best-fit value, the inaccuracy of the line lists, or the insufficiency of the wavelength calibration in data reduction. The observed wavelengths are shifted according to the inferred radial velocity, which are used in the following steps of our analysis whenever the observed and model spectra are compared together.

On occasion, the line parameters such as oscillator strength log($gf$) drawn from some line databases are not accurate enough or updated using more recent studies to well reproduce some specific observed spectral lines. To inspect the atomic line list, we have compared the log($gf$) of all identified atomic lines in the spectra of our target  from the Vienna Atomic Line Database (VALD3, \citealt{Ryabchikova2015,  Pakhomov2017,  Pakhomov2019}) with those from \textit{Linemake}\footnote{https://github.com/vmplacco/linemake} (an open-source atomic and molecular line list generator, \citealt{Placco2021}). We have found 11 lines that have different values of log($gf$) in these two line databases: Ti I (15334.85 {\AA}), Ti I (15602.84 {\AA}), Ti I (15715.57 {\AA}), Mg I (15748.99 {\AA}), Ca I (16197.07 {\AA}), Ca I (19853.09 {\AA}), Ca I (19933.73 {\AA}), Sc I (22052.14 {\AA}), Sc I (22065.23 {\AA}), Sc I (22266.73 {\AA}), and Ca I (22651.18 {\AA}). We have noted that only the three Ti I lines show better consistency between the observed spectrum and (Model){$_{\textrm{approx}}$ if log($gf$) values from \textit{Linemake}, rather than from the VALD3, are used. We have accordingly updated the log($gf$) of these three lines in the VALD3 line list using the values from \textit{Linemake} that are originally from \citet{lawler2013}. We have also replaced the FeH line list in our previous set used in Paper I \citep{Dulick2003} with the more recent one from \citet{Hargreaves2010}. This new line list reproduces synthetic models that are in significantly better agreement with observed spectra over regions dominated by FeH lines.

\section{Spectral Line Selection and Continuum/Pseudocontinuum Normalization}  \label{sec:line_normalize}
For our spectral fitting analysis, the ideal atomic and molecular lines are those that show consistency between the observed spectrum and the best-fit model. Since the best-fit model is undetermined before performing the fitting code, we compare the observed spectrum with (Model){$_{\textrm{approx}}$ over the spectral line candidates and select the best lines for the analysis. However, for a reasonable comparison, the observed spectrum needs to be locally continuum-normalized, or pseudo continuum-normalized for most regions where the flux level is lower than unity due to a large number of H$_{2}$O lines and the continuum cannot be identified. The reliability of inferred abundances therefore strongly depends on the propriety of spectral line selection, which relies on  the accuracy of the normalization process. 

Prior to normalizing the observed spectrum, the synthetic spectra are smoothed at the observed spectral resolution using a Gaussian kernel  and then interpolated at the shifted observed wavelengths. The continuum/pseudocontinuum normalization is performed using the method described in Paper I, which is based on the general concept presented in \citet{Santos-Peral2020}, but with some modifications. The most appropriate data points on the continuum/pseudocontinuum  around the analyzed spectral lines are chosen using a routine that implements a linear fit to the residuals R=O/S, where O is the observed flux and S is the synthetic flux, both at shifted, observed wavelengths, followed by two or three iterative {$\sigma$}-clippings. The value of the clippings changes from the first to the third iteration as 2{$\sigma$}, 1.5{$\sigma$}, and 1{$\sigma$}. For the cases where the three {$\sigma$}-clippings do not end up with enough number of normalizing data points, only the first two {$\sigma$}-clippings are performed.  The normalized spectrum is obtained after dividing the observed spectrum by the linear fit to the residuals of the final data points.

We identify the well-defined and almost isolated spectral lines that have a proper shape (e.g., not deformed by noise or bad pixels) and that are strong enough to be distinguished from the prevalent background molecular opacities (while these lines might still be weakly blended with molecular lines). We then look for the continuum/pseudocontinuum regions on both sides around these line candidates.  Often, a few lines are close together, and common normalizing regions  around,  and  in some cases, also between these lines, are determined. We test different pairs of continuum/pseudocontinuum ranges for each studied line (or a few lines if they are close together) and then normalize the observed spectrum using the process described above. We choose the pair of continuum/pseudocontinuum regions that lead to a normalized spectrum consistent with (Model){$_{\textrm{approx}}$ within those  regions. It should be noted that at least two normalizing data points (one each side)  are required to perform the final linear fit and normalize the observed spectrum. 

We further test the selected pairs of ranges by changing the corresponding elemental abundances and checking whether the normalizing points still remain on the continuum/pseudocontinuum. This inspection is important because, in the model fitting procedure, the observed spectrum is normalized relative to a number of model spectra with varying abundances before calculating {\ensuremath{{\chi}^2}} values (see Section \ref{sec:autospecfit}). As the abundance of an element varies while the physical parameters remain unchanged, the flux may be slightly changed over the neighboring regions around or even beyond of the respective spectral lines.  This flux redistribution could reshape the pseudocontinuum levels around some spectral lines and  even cause some weak absorption lines to appear. For example, after increasing the abundance of oxygen (that is linked to the abundance of H$_{2}$O), some H$_{2}$O absorption lines may arise within pseudocontinuum regions. In this case, the determined normalizing data points may show up inside the absorption lines that emerge after changing elemental abundances. 

We illustrate the normalizing regions and normalizing data points for a few spectral lines using the spectrum of our target in Figures \ref{fig:normalize_K_OH} and \ref{fig:normalize_Ti_Sc}.  Figure \ref{fig:normalize_K_OH} shows the synthetic flux of the normalizing data points (black circles at the edges of the panels) within the selected pseudocontinuum ranges on both sides of the  K I 15168.38 {\AA} line (left panels) and the OH 16526.25 {\AA} (right panels), which are separated from the inner spectral regions by green dashed-lines. The observed spectrum (red lines and circles) is normalized relative to (Model){$_{\textrm{approx}}$ (blue lines) as shown in the middle panels. The top and bottom left panels present the observed spectrum that is normalized relative to the model spectra similar to (Model){$_{\textrm{approx}}$, but with the relative abundance of potassium equal to [K/Fe]=$-$0.20 dex and [K/Fe]=+0.20 dex (or [K/H]=$-$0.03 dex and [K/H]=+0.37 dex, following the equation [X/H]=[X/Fe]+[M/H]), respectively. In the same way, the top and bottom right panels demonstrate the observed spectrum normalized with respect to the synthetic spectra similar to (Model){$_{\textrm{approx}}$, except for the relative oxygen abundance that is equal to [O/Fe]=$-$0.20 dex and [O/Fe]=+0.20 dex (or [O/H]=$-$0.03 dex and [O/H]=+0.37 dex), respectively. For both lines, although there is a slight change in the overall flux level with abundance variation, the shape of the pseudocontinuum in the selected ranges does not change, and the already chosen data points remain the most suitable normalizing points. If this was not the case, we would explore other ranges on the pseudocontinuum to find those that would meet the above condition.

Figure \ref{fig:normalize_Ti_Sc}} shows the same plots as Figure \ref{fig:normalize_K_OH}, but for two neighboring spectral lines, Sc I 22266.73 {\AA}  and Ti I 22274.02 {\AA}. Clearly, there is no observed pseudocontinuum region between these two lines that is in agreement with the synthetic models, and we, therefore, determine common normalizing ranges around both sides of the lines. The middle panels again display the observed spectrum normalized relative to (Model){$_{\textrm{approx}}$ (the two middle panels are repeated for better comparison between different abundances of each line, from top to bottom). In the top and bottom left panels, the relative abundance of Sc changes in the same way as in Figure \ref{fig:normalize_K_OH}, while the relative abundance of Ti is fixed, equal to the solar value. Similarly, in the top and bottom right panels, the relative abundance of Ti varies in the same manner as above, but the relative abundance of Sc is constant, equal to the solar value. For the two cases, the chosen normalizing ranges and data points  remain on the pseudocontinuum level and continue to be the best for the analysis. 

We examine the atomic and molecular (CO, OH, and FeH) line candidates and identify their best normalizing ranges, while adjusting for radial velocity if needed. We  then normalize the observed spectrum  according  to the synthetic spectra with  different relative abundances spanning from [X/Fe]=$-$0.30 dex to [X/Fe]=+0.30 dex with steps of 0.10 dex\footnote{We find this range of abundances sufficiently broad to examine all the studied lines and determine the best-fit model.} for the lines associated with the element X. We  visually compare the resulting normalized observed spectrum with the respective models, which gives us an early understanding of how consistently the synthetic models can reproduce the observed spectral lines assuming 1D LTE. For example, the two alkali lines,  K I 15163.07 {\AA} and  Na I 22056.4 {\AA}, show no adequate agreement with model spectra, which may be due to the insufficiency in the line lists, the NLTE effect \citep[]{Olander2021}, or other factors. For this reason, we removed these two lines from our analysis.

After a careful examination, we selected 148 spectral lines for 10 different species, CO, OH, Na, Mg, Al, K, Ca, Sc, Ti, and FeH, as listed in Table \ref{tab:line_data}. We then manually determine  a fitting or {\ensuremath{{\chi}^2}} window for the selected lines (the third column of Table \ref{tab:line_data}), mainly around the cores and far from the outermost part of the wings (that are more influenced by spectral noise), to perform the {\ensuremath{{\chi}^2}} minimization. For some adjoining doublet lines of the same species (e.g., some OH lines), a single common {\ensuremath{{\chi}^2}} window is defined. We use the adopted normalizing ranges and {\ensuremath{{\chi}^2}} windows of the  lines as input to run AutoSpecFit in the next step. 

As presented in the last column of Table \ref{tab:line_data}, four atomic lines in our line set (Na I 22083.66 {\AA}, Al I 16718.96 {\AA}, Al I 16750.56 {\AA}, and Sc I 22266.73 {\AA})  are a combination of multiple lines, including lines from hyperfine structure (HFS) splitting. HFS data have been included to the  VALD3 database \citep{Pakhomov2017, Pakhomov2019}, and have shown to properly model the HFS lines in M dwarfs  \citep{Shan2021}. In addition, several  atomic lines (Mg I 15765.84 {\AA}, Ti I 15334.85 {\AA}, Ti I 15715.57 {\AA}, Ti I 21782.94 {\AA}, Ti I  22211.24 {\AA}, Ti I  23441.48 {\AA}) are blended with  a few other lines associated with the same elements, most of which are  too weak to influence the shape of the main (central) lines (Table \ref{tab:line_data}). It should be pointed out that  every single line is included in the line list and modeled by our spectral synthesis.

\section{Physical Parameters of Planet-Host M dwarf K2-18}\label{sec:all-k2-18-par}

\subsection{Effective Temperature, Metallicity, and Surface Gravity}\label{sec:k2-18-par}

K2-18, an M2.5V dwarf star (\citealt{Schweitzer2019}), resides in the Solar vicinity at a distance of 38 pc (\citealt{GaiaDR32021}). Due to its proximity, numerous studies in the literature have determined its stellar parameters. These studies report an effective temperature of approximately 3500 K (\citealt{Montet2015}, \citealt{Stassun2019}, \citealt{Martinez2017}, \citealt{Schweitzer2019}, \citealt{Zhang2020ApJS}, \citealt{Reiners2022}), surface gravity around 4.70 (\citealt{Stassun2019}, \citealt{Schweitzer2019}, \citealt{Shan2021}, \citealt{Queiroz2023}), and metallicity varying notably across different works, from $-$0.30 (\citealt{Ding2022}) to +0.26 dex (\citealt{Hardegree2020}).

We use our $H-$band spectra to determine the atmospheric parameters of K2-18 ($T_{\rm eff}$ and log $g$) using the methodology of \cite{Souto2020}. To summarize, we derive oxygen abundances from two species (H$_{2}$O and OH lines) for a set of effective temperatures ranging from 3200 to 3800 K in steps of 100 K. Because the H$_{2}$O and OH lines display different sensitivity to changes in $T_{\rm eff}$, there will be a unique solution yielding the oxygen abundance to a $T_{\rm eff}$ value. To derive the surface gravity, we employ the same methodology but determine the oxygen abundance for a set of log($g$) from 4.50 to 5.00 in steps of 0.10 dex. The abundances are inferred from the best-fit synthetic models compared to the observed spectrum that was generated in the same way as described in this study, i.e., employing the Turbospectrum code in conjunction with MARCS models, as well as using  the APOGEE line list (\citealt{Smith2021}).  To derive the uncertainties in the atmospheric parameters ($T_{\rm eff}$ and log $g$), we propagate the oxygen abundance uncertainty into the atmospheric parameter determination methodology. For K2-18,  we obtain $T_{\rm eff}$ = 3547 $\pm$ 85 and log $g$ = 4.9 $\pm$ 0.10. Another product from this analysis is the stellar overall metallicity, which is determined from the Fe I and FeH lines available in the $H-$band (see \citealt{Souto2020}, \citealt{Souto2021}). We obtain that K2-18 is metal-rich, where [Fe/H] = +0.17 $\pm$ 0.10 dex.  We adopt the uncertainty of metallicity from \cite{Melo2024}.

It is important to emphasize that all steps of the parameter determination procedure are completely different from our abundance analysis described in this paper. Nevertheless, we find very good agreement between  (Model){$_{\textrm{approx}}$ associated with the derived physical parameters and the observed spectrum, which assures the reliability of the abundances based on these parameters.

\subsection{Microturbulent Velocity}  \label{sec:vmic}
Low-mass stars generally have microturbulence between 0 and 2 km/s \citep[e.g.][]{Reid_Hawley2005, Bean2006,  Tsuji_Nakajima2014, Pavlenko2017, Souto2017, Olander2021, Recio-Blanco2023}. We determine the microturbulence $\xi$ using the approach described in Paper I. We start with the three species having the most abundant selected lines, i.e., CO (whose abundance is an indicator of carbon abundance), OH (whose abundance is an indicator of oxygen abundance), and FeH (whose abundance is an indicator of iron abundance).  We use the molecular FeH lines to measure iron abundance because they are significantly more numerous than atomic Fe I lines. It is important to note that there is a difference between the iron abundance inferred from the methodology of \citealt{Souto2020} and \citealt{Souto2021} (Section \ref{sec:k2-18-par}) and that derived from this analysis (Section \ref{sec:application}), though consistent within the errors. However, the former abundance has shown to be a very good estimate of the overall metallicity of the target and has been used to measure the abundances of the analyzed elements. 

For any of these three species, we generate a grid of models, all associated with the target's parameters but having different values of $\xi$ ranging from 0 to 2 km/s with steps of 0.1 km/s, and different relative abundances  spanning from [X/Fe]=-0.30 dex to [X/Fe]=+0.30 dex with steps of 0.01 dex, where X denotes C, O, or Fe, while assuming the solar relative abundance for all other elements Y (([Y/Fe]=0), leading to 1281 synthetic spectra in total for each species. We then perform a single {\ensuremath{{\chi}^2}} minimization routine (Section \ref{sec:autospecfit}) over all the spectral lines corresponding to each of the three aforementioned molecules individually. To this end, the observed lines  are normalized  relative to the model spectra that correspond to a specific value of $\xi$ and varied abundances of the respective element and then compared to those models to obtain the best-fit abundance. We calculate the average and the standard deviation of abundances for each species and each $\xi$ value. We find the CO lines the most sensitive to $\xi$, as the average of CO abundances shows the largest variation as a function of $\xi$. The standard deviation of CO abundances is minimum when $\xi$=1 km/s, and we therefore adopt $\xi$=1.0 $\pm$ 0.1 km/s for K2-18 in our analysis. As we see in Section \ref{sec:application}, the CO spectral lines are indeed the most sensitive to microturbulent velocity as compared to the lines of all other studied species (Table \ref{tab:results}). It should be noted that the effect of  rotational velocity and magnetic field on the target's spectrum is negligible, and we do not include these two physical parameters in our study.

\subsection{Mass, Radius, and Age} \label{sec:age-mass-radius}
Since  M-dwarf seismology is infeasible \citep{Rodriguez-Lopez2019}, the mass and age of M dwarfs are determined using some indirect techniques. For example,   \cite{Guinan-Engle2019} estimated an age of $\sim$2.4$\pm$0.6 Gyr for K2-18 using the P{$_{rot}$}-Age relation  for M2.5–6.0 stars \citep{Engle-Guinan2018}, as K2-18 has a well-determined rotation-period P{$_{rot}$} = 39.6 $\pm$ 0.9 days \citep{Sarkis2018}. One may consider a theoretical method using stellar isochrones and evolution tracks to infer M-dwarf age. However, M dwarfs evolve very slowly once they reach the main sequence and there is no age dependence associated with these methods.

M-dwarf Mass can be estimated using mass-luminosity relation (MLR, e.g., \citealt{Benedict2016, Mann2019}) that connect the luminosity of a lower-main-sequence star to its mass. \cite{Cloutier2019} derived a mass of  0.495 $\pm$
0.004 M$_{\sun}$ for the host star K2-18 using the MLR from \cite{Benedict2016} based on absolute K-band magnitudes (which is favored over the V-band whose dispersion about the relation
is twice that in the K-band.)

Interferometry can be used to accurately measure the angular diameter, which together with a well-measured bolometric flux can yield an accurate $T_{\rm eff}$ measurement. However, this technique is expensive in terms of time and analysis, and limited to stars that are sufficiently large ($\gtrsim$0.3 mas) and bright ($\gtrsim$8 mag). Empirical relations are therefore more appropriate to derive M-dwarf radius. For instance, \cite{Cloutier2019} estimated a radius of  0.469 $\pm$ 0.010 R$_{\sun}$ for our target K2-18 using  the mass-radius relationship for M dwarfs from  \cite{Boyajian2012}.

The above inferred  mass and radius of K2-18 from empirical relations  are not  accurate enough to improve our inferred values of  $T_{\rm eff}$ and log $g$ using high-resolution spectroscopy. The advantage of our method is that these two parameters can be derived consistently from the same spectra using the same diagnostic features. This is possible, thanks to the excellent quality of our IGRINS spectra, which allows deriving consistent parameters for similar stars observed with the same instrument.

\section{AutoSpecFit: An Automatic Model Fitting Code for Elemental Abundance Measurements}  \label{sec:autospecfit}
We present the AutoSpecFit code that carries out an automatic line-by-line {\ensuremath{{\chi}^2}} minimization in an iterative manner and allows Turbospectrum to generate the required synthetic spectra for each iteration without interrupting the run. The abundances of the selected lines are determined separately and modified in each iteration until the final abundances are reached.  The selected normalizing ranges and {\ensuremath{{\chi}^2}} windows are used as input for running the code. Physical parameters, i.e., effective temperature $\rm{T_{eff}}$,  metallicity [M/H], surface gravity log($g$), and microturbulent velocity $\xi$) are also required in advance to execute AutoSpecFit, and these parameters are not changed during the run. We find the spectral lines sensitive to variations in physical parameters, and as a result, these parameter can be degenerate with chemical abundances, causing significant uncertainties in inferred abundance values. We accordingly use the derived parameters from other independent methods (see Section \ref{sec:k2-18-par} and also Section \ref{sec:vmic} for the  microturbulence parameter inferred from an examination independent of AutoSpecFit) and keep them fixed with no further adjustment when measuring elemental abundances.

The pipeline first generates a number of synthetic spectra for each studied element X associated with the physical parameters of the star ($\rm{T_{eff}}$}, [M/H], log($g$), and $\xi$), but varied relative abundances of that particular element usually ranging from [X/Fe]=$-$0.30 dex to [X/Fe]=+0.30 dex in steps of 0.01 dex (61 models for each element) that are needed for a detailed abundance analysis, and the solar relative abundances ([Y/Fe]=0) for all other elements Y. These spectra are used in the first iteration of {\ensuremath{{\chi}^2}} minimization as follows. The observed spectrum is normalized relative to all the synthetic models over each spectral line. We perform the normalization process during each iteration, i.e., normalizing the observed spectrum with respect to each model under examination before calculating {\ensuremath{{\chi}^2}}. This is in contrast with some other studies in which the observed spectrum is normalized relative to a first-guess model spectrum and then used in the {\ensuremath{{\chi}^2}} minimization routine without any further change \citep[e.g.][]{Kirby2010, Sarmento2021, Recio-Blanco2023}. However, it is important to note that the variation of abundances generally results in a change in the flux level of model spectra. For example, the right panels of Figure \ref{fig:normalize_K_OH} show a noticeable shift in the overall flux level of the models around the OH line by changing the relative abundance of oxygen from [O/Fe]=$-$0.20 dex to [O/Fe]=+0.20 dex.  Since the observed spectrum is normalized relative to each of these three models, it is also scaled in the same way as the models, and a proper comparison can thus be made between the observed spectrum and the models for different abundances. This is the reason why we prefer to normalize the observed spectrum relative to all the models used in each minimization to have a meaningful comparison.

The observed flux errors are also normalized with the same linear fit  used to normalize the observed spectrum. These normalized errors are then included in the {\ensuremath{{\chi}^2}} formula as below:

\begin{equation}
{\chi}^2 = \sum_{i}^{}\frac{\rm{(O_{i}-S_{i})^{2}}}{\rm {{(Oerr)_{i}}}^{2}}
\end{equation}

\noindent
where $\rm{O_{i}}$ is the continuum/pseudocontinuum-normalized, observed flux, $\rm{S_{i}}$ is the continuum-normalized, synthetic flux, and $\rm{Oerr_{i}}$ is the normalized, observed flux error (as described above), all at the observed, shifted wavelength ``i". The {\ensuremath{{\chi}^2}} value is calculated within the defined {\ensuremath{{\chi}^2}} window of each spectral line (Section \ref{sec:line_normalize}). Using the generated models, the {\ensuremath{{\chi}^2}} related to each model within the chosen {\ensuremath{{\chi}^2} or fitting window of any selected spectral line is calculated, and a polynomial fit is implemented to the resulting {\ensuremath{{\chi}^2}} values as a function of abundances. The abundance that minimizes the polynomial function is recorded as the best-fit abundance of each particular line. For those elements that have more than one spectral line, we calculate the average abundance following the approach described in \citet{Adibekyan2015}. We use a weighted mean  with the inverse of the distance from the median abundance as a weight, where the distance is expressed in terms of the standard deviation (SD) of the abundances. Since the weights corresponding to the lines with abundances close to the median abundance are very high, we bin the distances with a size of 0.1$\times$SD. In this way, a weight of 1/(0.1$\times$SD) is given to the lines with abundances that are between 0 and 0.1$\times$SD  away from the median abundance,  a weight of  1/(0.2$\times$SD) is given to the lines with abundances that are between 0.1$\times$SD and 0.2$\times$SD  away from the median abundance, and so on. We prefer this method, which reduces the impact of outlier lines without removing them. \citet{Adibekyan2015} argue that the detection of real outliers is a difficult task, and the commonly-used outlier-removal methods \citep[e.g.][]{Tukey1977, Shiffler1988, Iglewicz_Hoaglin1993, Carling2000} are dependent on the models and the applied thresholds, and also are not based on a clear prescription or a theoretical foundation. The authors, therefore, recommend the use of a weighted mean instead of any outlier-removal technique.

The abundance of elements with a single line or the average abundance of elements with multiple lines inferred from the first iteration is used for the second iteration. A number of model spectra are generated for each element X, again associated with the target's parameters and varied relative abundances of that specific element ranging from [X/Fe]=$-$0.30 dex to [X/Fe]=+0.30 dex in steps of 0.01 dex, but with relative abundances of all the other studied elements Y inferred from the first iteration\footnote{It should be noted that the relative abundance of the non-studied elements remain to be the solar values, which are the default abundances when running Turbospectrum without any abundance customization.}. These new synthetic spectra are used in the model fitting process exactly in the same way as the first iteration, and an average abundance for each element of interest is derived using the procedure as outlined above.  The algorithm is repeated, and every time a series of model spectra are generated that are optimized by the abundances obtained from the previous iteration and employed in the next one until the abundances converge to their final values, i.e.,  the difference in inferred abundance between two consecutive iterations is less than 0.01 dex. When this condition is met for all the studied elements simultaneously, the abundances are recorded as the final best-fit values, and the code stops. Figure \ref{fig:flowchart} shows a flowchart of the performance of AutoSpecFit.

AutoSpecFit allows Turbospectrum to automatically produce the model spectra required for each iteration ``on the fly". This is an advantage over traditional methods in which the models with all possible combinations of elemental abundances need to be generated in advance because the abundances obtained from each iteration are unknown prior to running the fitting code. However, for a detailed abundance measurement, this would lead to an extremely large number of model spectra. For example, in this study, the combinations of the 61 abundances for 10 elements would require  {$\ensuremath{61^{10} \simeq 7\times10^{17}}$} spectra with traditional grid sampling. The generation of this number of synthetic spectra would be computationally intensive and exceedingly time-consuming, even using high-performance computing systems, which is practically impossible. Instead, our pipeline produces {$\ensuremath{61 \times 10=610}$} models for each iteration, and for instance, an analysis with  15 iterations (which is more than enough for a typical abundance measurement, see Section \ref{sec:application}) would require 9150 models in total, which is computationally manageable to generate\footnote{We make use of a high-performance computing system which enables us to produce 610 model spectra within around 6 hours through 10 parallel jobs (corresponding to 10 elements). With an additional (less than) one hour for the fitting process (given that our original code is in MATLAB), each iteration takes around 7 hours, on average. For a typical analysis with 8 iterations, the total time to perform the AutoSpecFit is $\sim$56 hours or $\sim$2.3 days.}. 

In addition, AutoSpecFit enables us to take into account the complex impact of the abundance variation of different elements on each other.  A change in the abundance of an element (while the physical parameters are kept constant) may cause a slight flux redistribution over different regions, which can be reflected in the abundance measurements of other elements. That is why we use an iterative spectral fitting routine to account for this effect, which can be perceived by the abundance change of an element from one iteration to another (Figures \ref{fig:iteration} and \ref{fig:iteration_temp}-\ref{fig:iteration_vmic_neg}). The code proceeds until all elements reach their final abundances that are globally consistent.   

\section{Application of AutoSpecFit to the Planet-Host M dwarf K2-18}  \label{sec:application}

\subsection{Chemical Abundances}  \label{sec:abundances}
We apply our technique to the planet-host M dwarf K2-18 to measure the abundances of 10 elements: C (using CO lines), O (using OH lines), Na, Mg, Al, K, Ca, Sc, Ti, and Fe (using FeH lines), as listed in the first column of  Table \ref{tab:results}. The number of the lines corresponding to each species, N, is presented in the second column of this table. As already mentioned, the star's physical parameters, i.e., $T_{\rm eff}$ = 3547 $\pm$ 85 K, [M/H] = 0.17 $\pm$ 0.10 dex, log($g$) = 4.90 $\pm$ 0.10 dex, and $\xi$ = 1.0 $\pm$ 0.1 km/s, as well as the selected normalizing ranges and {\ensuremath{{\chi}^2}} windows are used as input to run the AutoSpecFit. The fitting process converges after five iterations. Figure \ref{fig:iteration} shows how the elemental abundances change from one iteration to another until reaching their final best values, which clearly indicates the correlation between the abundances of different elements.

The number of lines corresponding to each element is shown in the second column, and the resulting abundances ([X/H]) are shown in the third column of Table \ref{tab:results}. We obtain a carbon-to-oxygen ratio for our target C/O=0.568 (for reference, the solar ratio is (C/O)$_{\sun}$=0.540 using the solar abundances from \citet{Grevesse2007}). We also determined the abundance ratios associated with several planet-building elements such as Al/Mg=0.080, Ca/Mg=0.065, and Fe/Mg=0.698. Figure \ref{fig:best_fit} compares the normalized observed spectrum (red lines and circles) and the final best-fit model (blue lines) that corresponds to the target's parameters and the derived abundances over 10 spectral lines related to the 10 analyzed elements.

\subsection{Abundance Errors}  \label{sec:errors}
To determine the parameter sensitivity and the systematic uncertainties of the derived abundances, we deviate the physical parameters by their errors (Sections \ref{sec:k2-18-par} and \ref{sec:vmic}), in both positive and negative direction one at a time, i.e.,  $T_{\rm eff}$ + 85 = 3632 K, $T_{\rm eff}$ $-$ 85 = 3462 K, [M/H] + 0.10 = 0.27 dex, [M/H] $-$ 0.10 = 0.07 dex, log($g$) + 0.10 = 5.00 dex, log($g$) - 0.10 = 4.80 dex, $\xi$ + 0.10 = 1.10 km/s, and $\xi$ - 0.10 = 0.90 km/s. We then perform the AutoSpecFit code eight times, in each of which only one parameter is deviated while the other parameters remain the same as the target's parameter values, and the abundances of the analyzed elements are obtained from each run.  Using the synthetic models associated with the targets' parameters but only one parameter departed by its error, we visually inspect the normalizing ranges over the selected spectral lines and find these regions are still appropriate for normalizing observed spectrum even with abundance variation. This assures us, for our future studies, that once we determine the best normalizing ranges relative to the models with the target's parameters, they can also be used for models with parameters that are deviated by their errors. Large departures beyond typical parameter uncertainties would definitely require a new set of normalizing ranges.

Figures \ref{fig:iteration_temp}-\ref{fig:iteration_vmic_neg} in the Appendix display the abundance of the 10 studied elements as a function of iteration number for eight AutoSpecFit runs using different input parameters, as shown in the captions. The number of iterations required for performing the AutoSpecFit using the deviated parameters is generally equal or more than that required for running the code using the target's parameters (Figure \ref{fig:iteration}). For each case,  the abundances change more significantly in the first few iterations, and then smoothly converge towards their final values.

In Table \ref{tab:line_data}, the columns 4-11 show the abundance variation due to the deviated parameters relative to the abundances obtained from the models with the star's parameters. The 
abundance variation of each element depends on the deviated parameter, as elemental abundances
 show different sensitivities to different parameters as well as the direction these parameters change. In addition, the abundance variation differs from one element to another for the same parameter change. For example, the abundance of the elements Ca, Al, and Mg are most sensitive to $T_{\rm eff}$, while the abundance of the light element C (from CO lines) is least sensitive to $T_{\rm eff}$. The abundance of the element Na shows the highest sensitivity to [M/H], but the abundance of the elements C, O, K, and Sc shows no significant sensitivity to [M/H].  The abundances of all the 10 studied elements are rather sensitive to log($g$), with the elements Al and Ca having the highest and K having the lowest sensitivity. The variation of microturbulence velocity $\xi$ generally has a weaker influence on the elemental abundances compared to other parameters \citep[e.g.][]{Souto2022, Hejazi2023}, with the abundance of element C showing the highest sensitivity to  $\xi$.
 
 We take an average of the absolute values of the two abundance variations related to the change of each  parameter in two directions (i.e., negative and positive). We then calculate the quadrature sum of these four averages for each element as the systematic abundance error, $\rm{\sigma_{sys}}$, which is shown in the column 12 of Table \ref{tab:results}. We also obtain the random (statistical) abundance error of the four species, CO, OH, Ti, and FeH that have a statistically large  number of lines, i.e., N $\geq$ 10, using the standard  error of the mean, i.e., $\rm{\sigma_{ran}}$=std/$\rm{\sqrt{N}}$, where std is the standard deviation of  the abundances from different  lines of each species, as shown in the column 13 of  Table \ref{tab:results}. The last column of the table presents the quadrature sum of the systematic and random (if applicable) errors, as the total error of the derived abundances. It should be noted that random errors are too small to significantly contribute to the total errors. For those elements with no random error, the total error may be slightly underestimated.

Figure \ref{fig:abundance_atomicnumber} presents the abundances of the 10 analyzed elements as a function of  their atomic number, and their total abundance errors are shown as vertical error bars. Using the abundance errors, we obtain the uncertainty of the abundance ratios: C/O=0.568 $\pm$ 0.026,  Al/Mg=0.080 $\pm$ 0.011, Ca/Mg=0.065 $\pm$ 0.010, and Fe/Mg=0.698 $\pm$ 0.178. We recall that the abundance ratio of two elements depends on the subtraction of their absolute abundances, i.e., $\rm{X/Y=10^{(A(X)-A(Y))}}$, and as a result, their systematic uncertainties  related to the variation of different stellar parameters  largely cancel. In addition, the (uncorrelated) random uncertainties (if applicable) are very small (see Table \ref{tab:results}). All these have led to relatively small uncertainties of abundance ratios, other than Fe/Mg for which the rather large difference between the systematic errors of the two elements Fe and Mg associated with effective temperature has resulted in a significantly larger uncertainty.

It is important to note that abundance errors highly depend on the uncertainty of input physical parameters. Smaller deviations of parameters, in particular effective temperature,  would give rise to smaller abundance errors \citep{Melo2024}.  To derive more accurate elemental abundances,  we need to have more accurate input stellar parameters, which requires more reliable model atmospheres and line lists as well as more robust techniques for parameter determination.

\begin{figure*}[hbt!]
\centering
\begin{subfigure}
  \centering
  \includegraphics[width=0.518\linewidth]{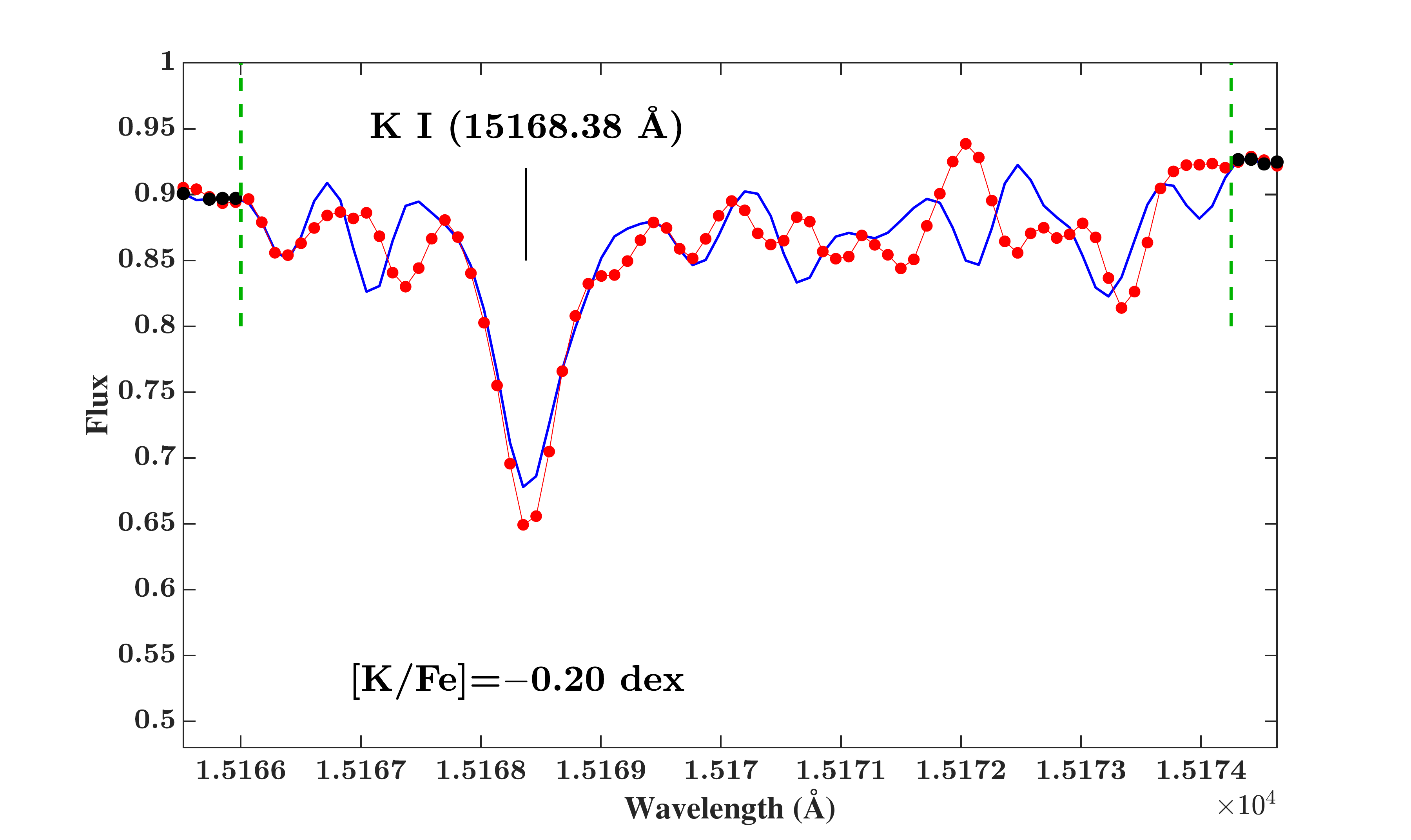}
\end{subfigure}
\hspace{-1.1cm}
\begin{subfigure}
  \centering
  \includegraphics[width=0.518\linewidth]{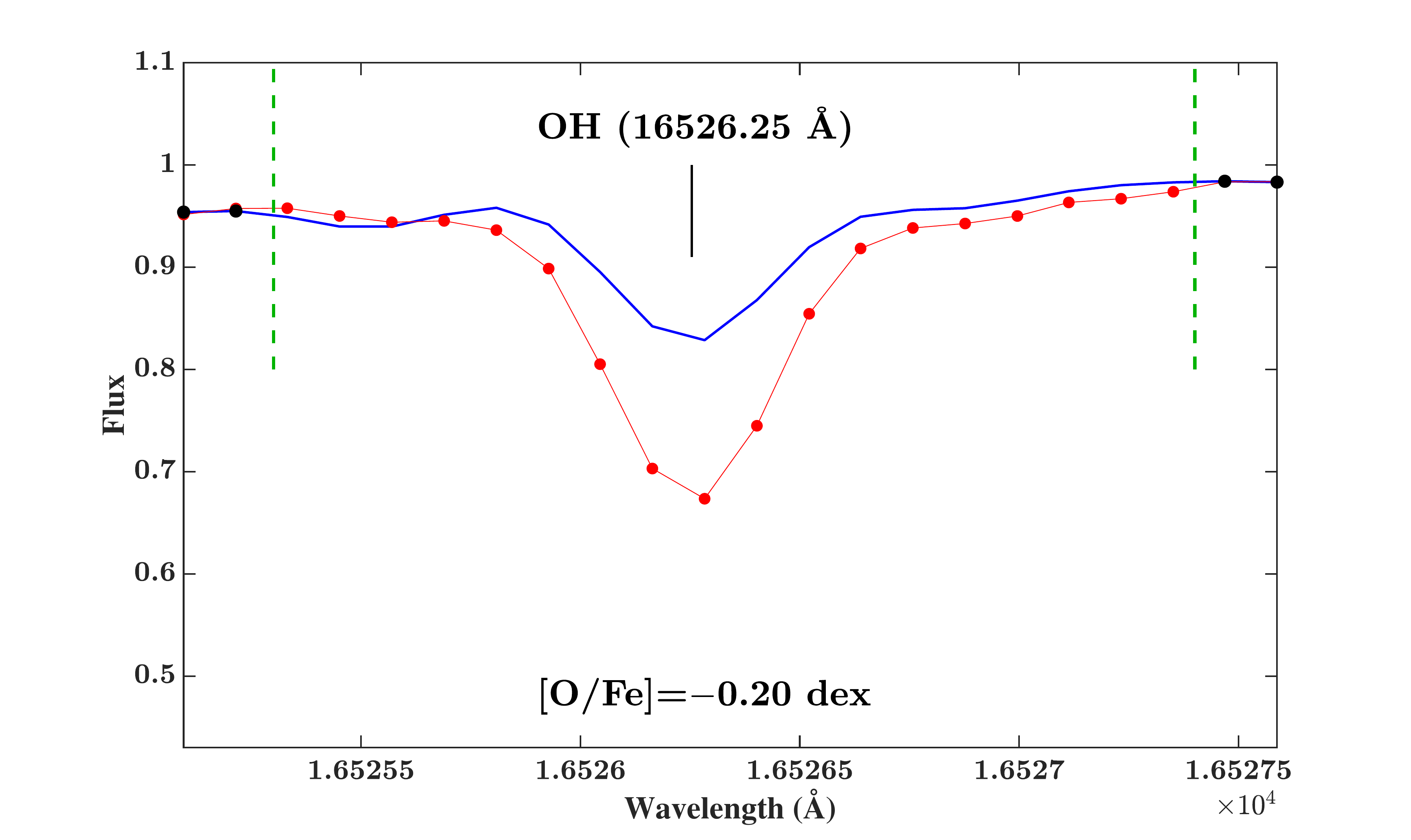}
\end{subfigure}
\vfill
\begin{subfigure}
  \centering
  \includegraphics[width=0.518\linewidth]{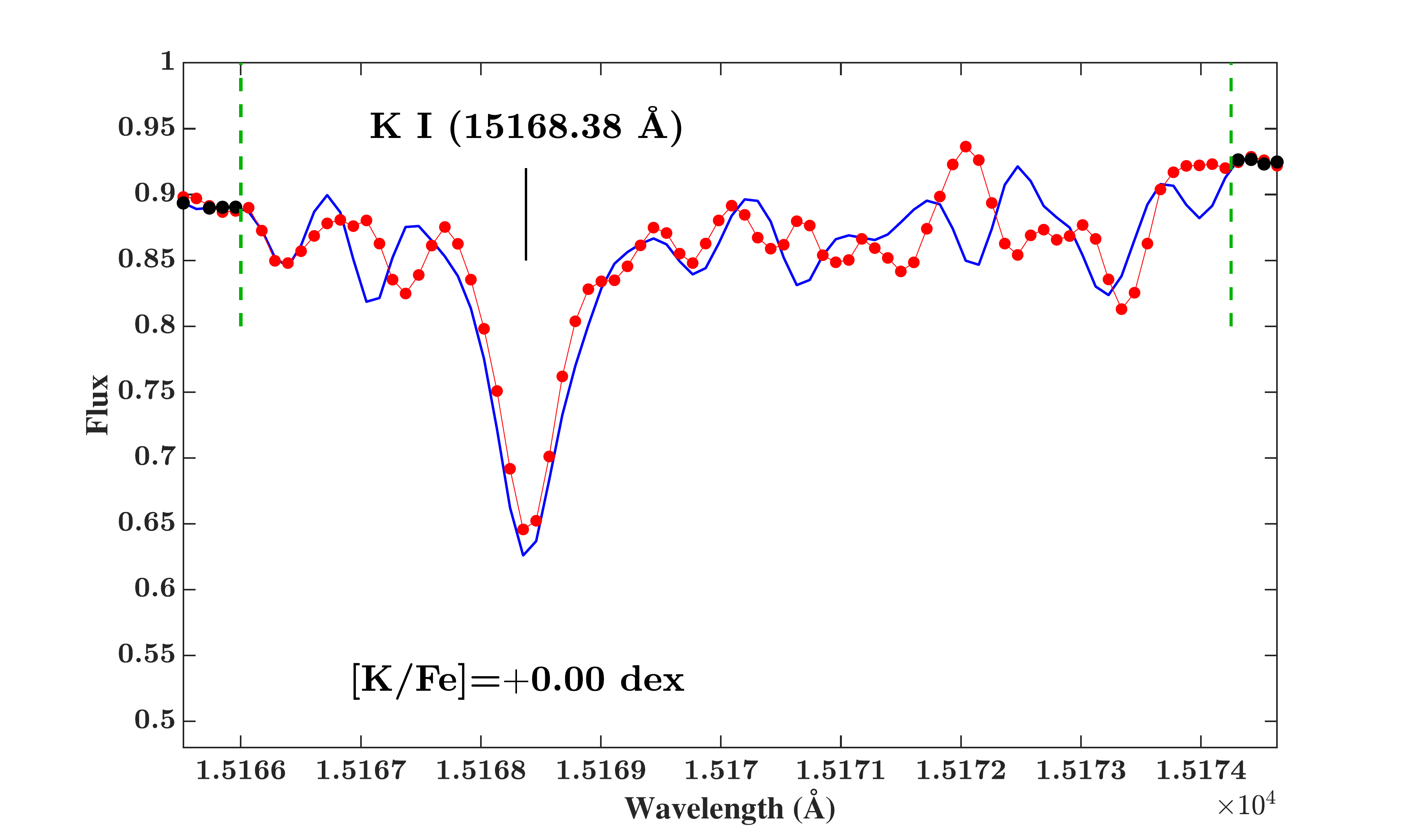}
\end{subfigure}
\hspace{-1.1cm}
\begin{subfigure}
  \centering
  \includegraphics[width=0.518\linewidth]{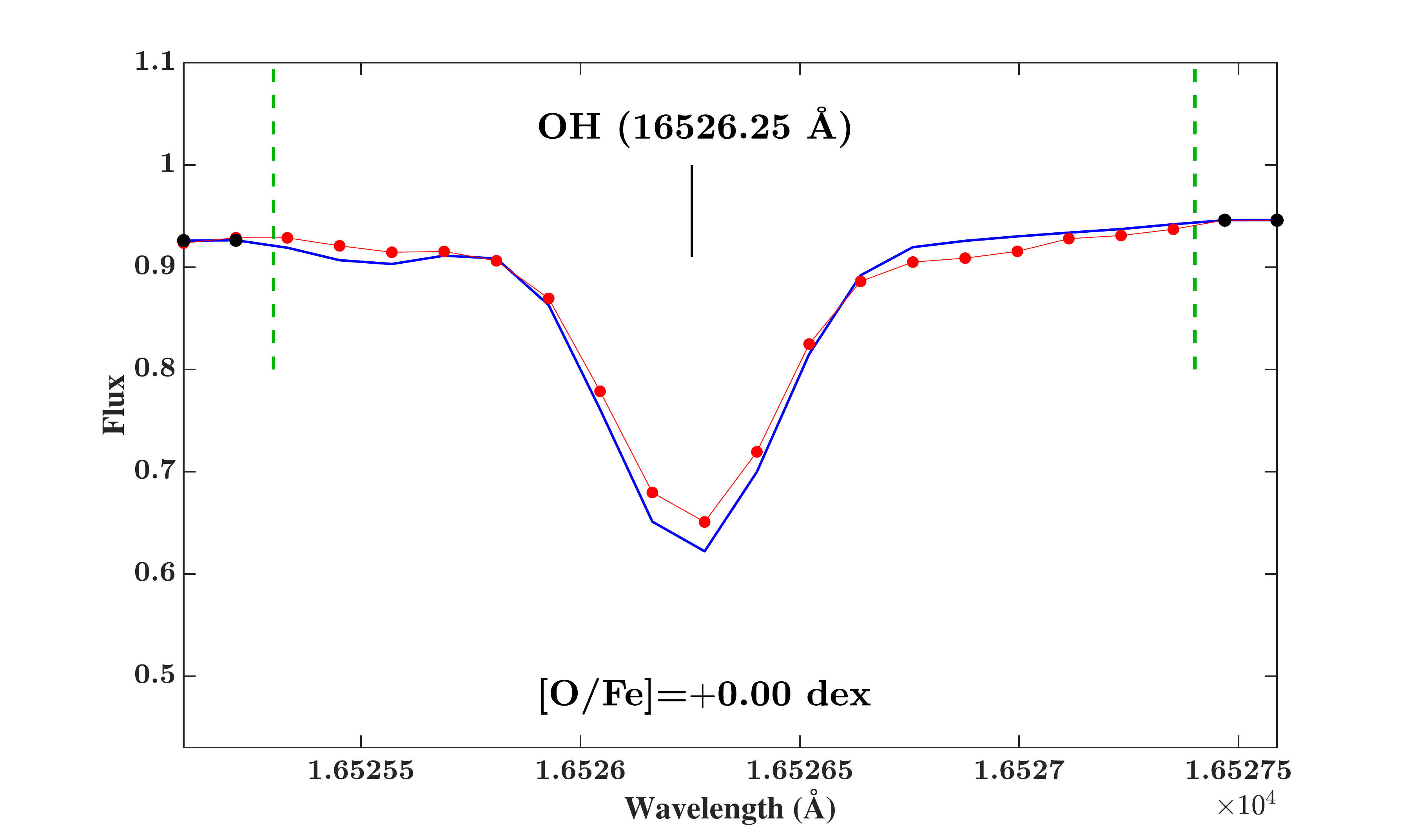}
\end{subfigure}
\vfill
\begin{subfigure}
  \centering
  \includegraphics[width=0.518\linewidth]{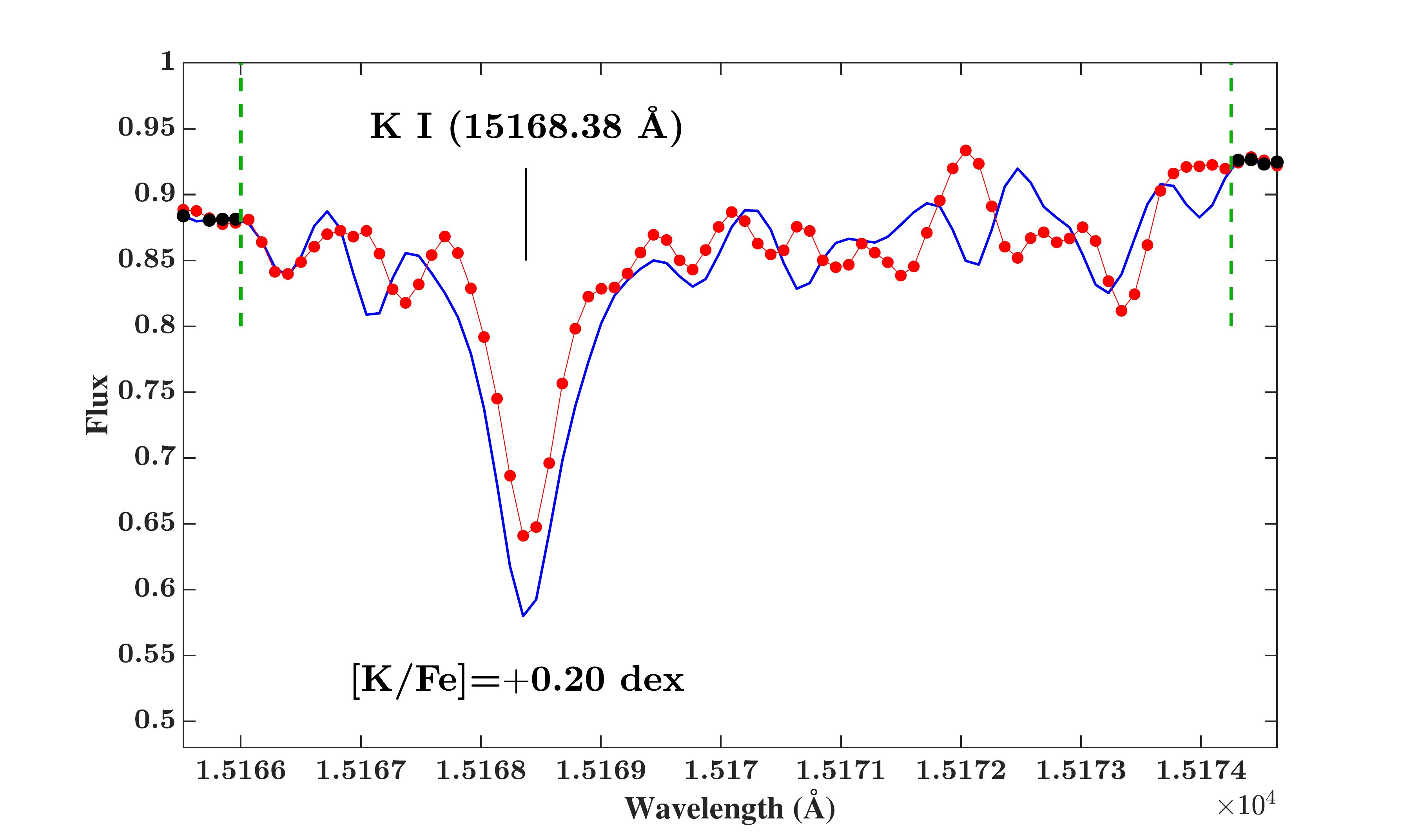}
\end{subfigure}
\hspace{-1.1cm}
\begin{subfigure}
  \centering
  \includegraphics[width=0.518\linewidth]{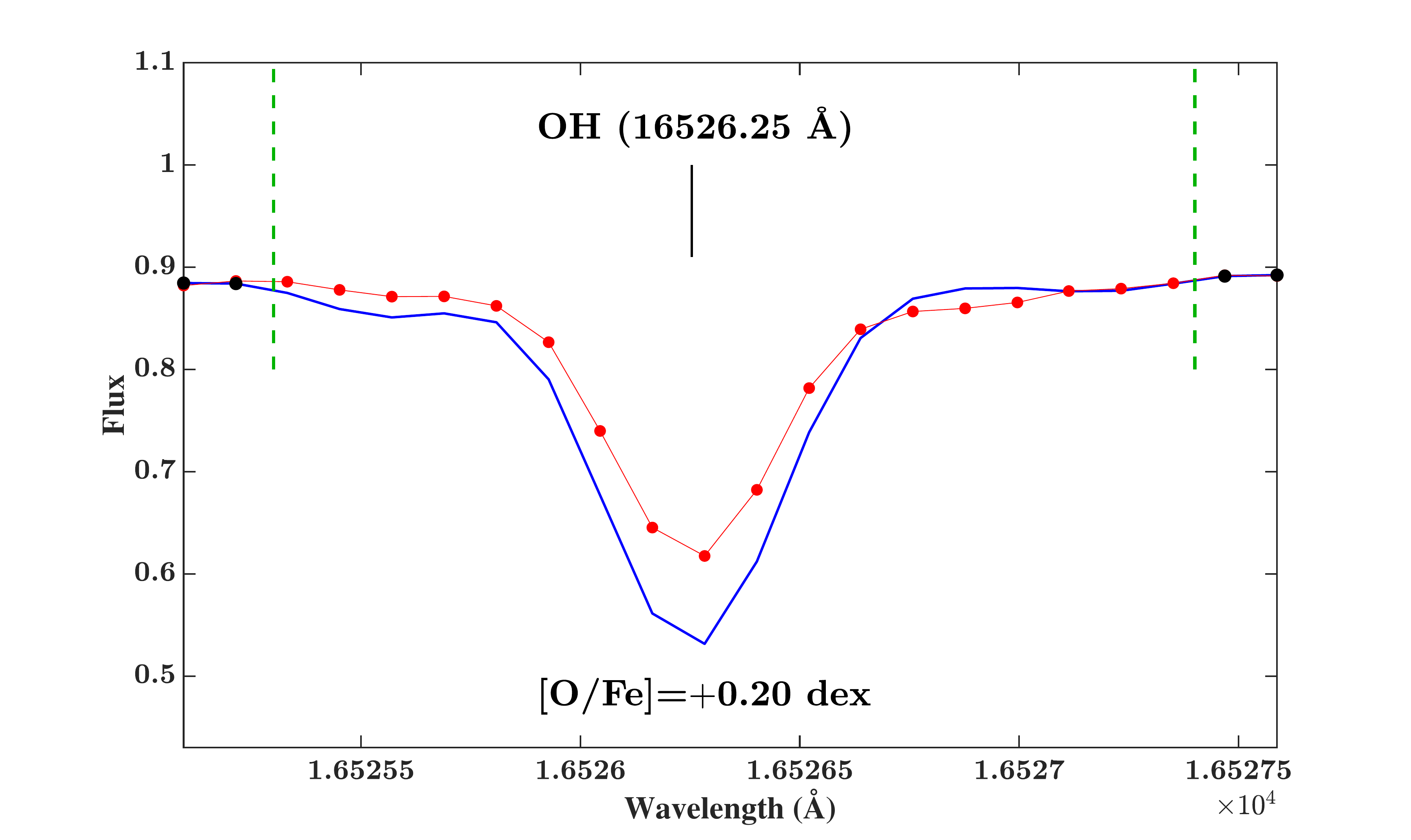}
\end{subfigure}
\caption{Comparison between the normalized observed spectrum (red lines and circles) of K2-18 and the model spectra (blue lines) associated with the target's parameters but varying abundances of the element K  (left panels) and the element O (right panels), while assuming solar relative abundances for all other elements. The black circles (at the edges of the panels) show the normalizing points within the selected continuum/pseudocontinuum normalizing ranges that are separated from the inner spectral regions by green dashed lines.}
\label{fig:normalize_K_OH}
\end{figure*}

\begin{figure*}[hbt!]
\centering
\begin{subfigure}
  \centering
  \includegraphics[width=0.518\linewidth]{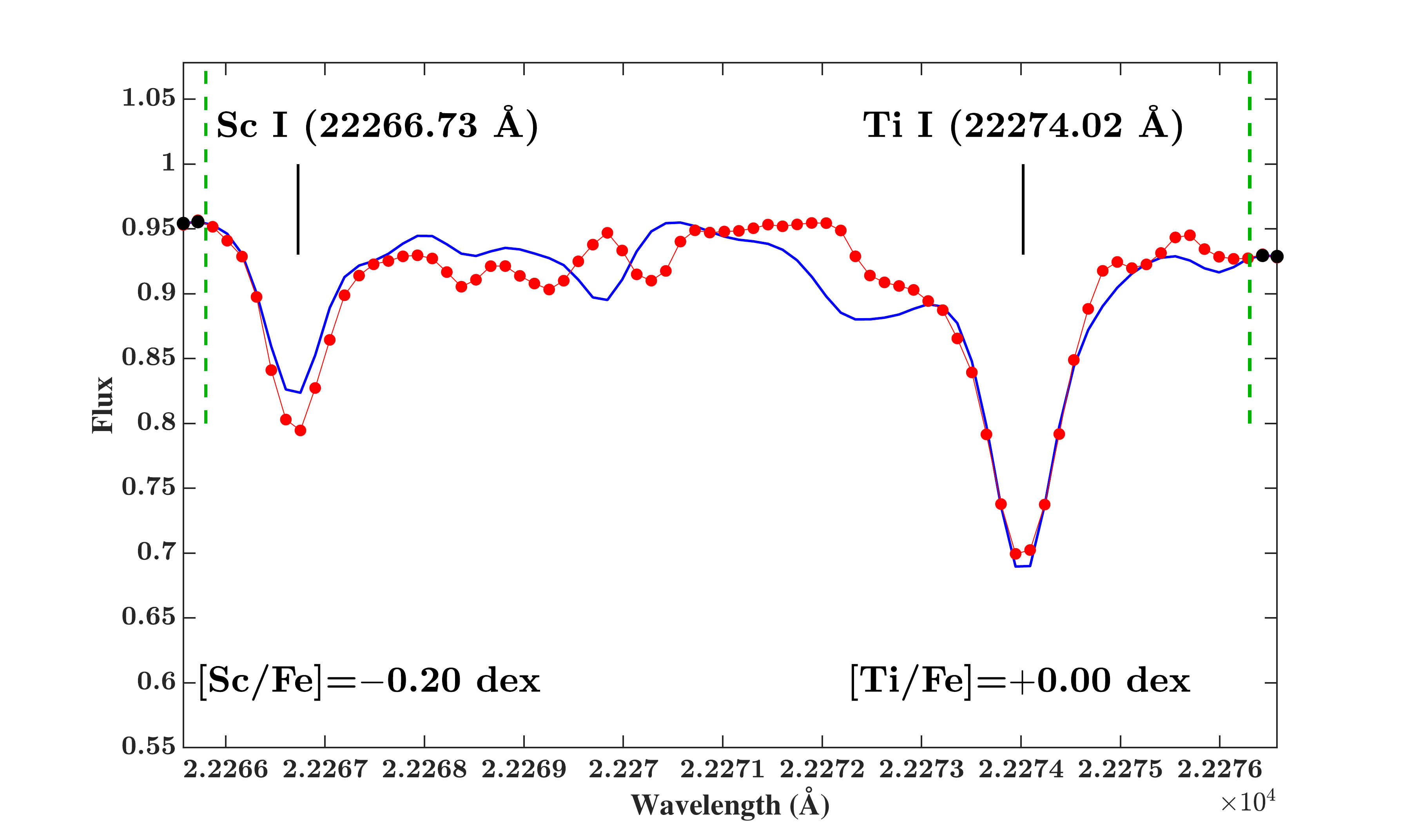}
\end{subfigure}
\hspace{-1.1cm}
\begin{subfigure}
  \centering
  \includegraphics[width=0.518\linewidth]{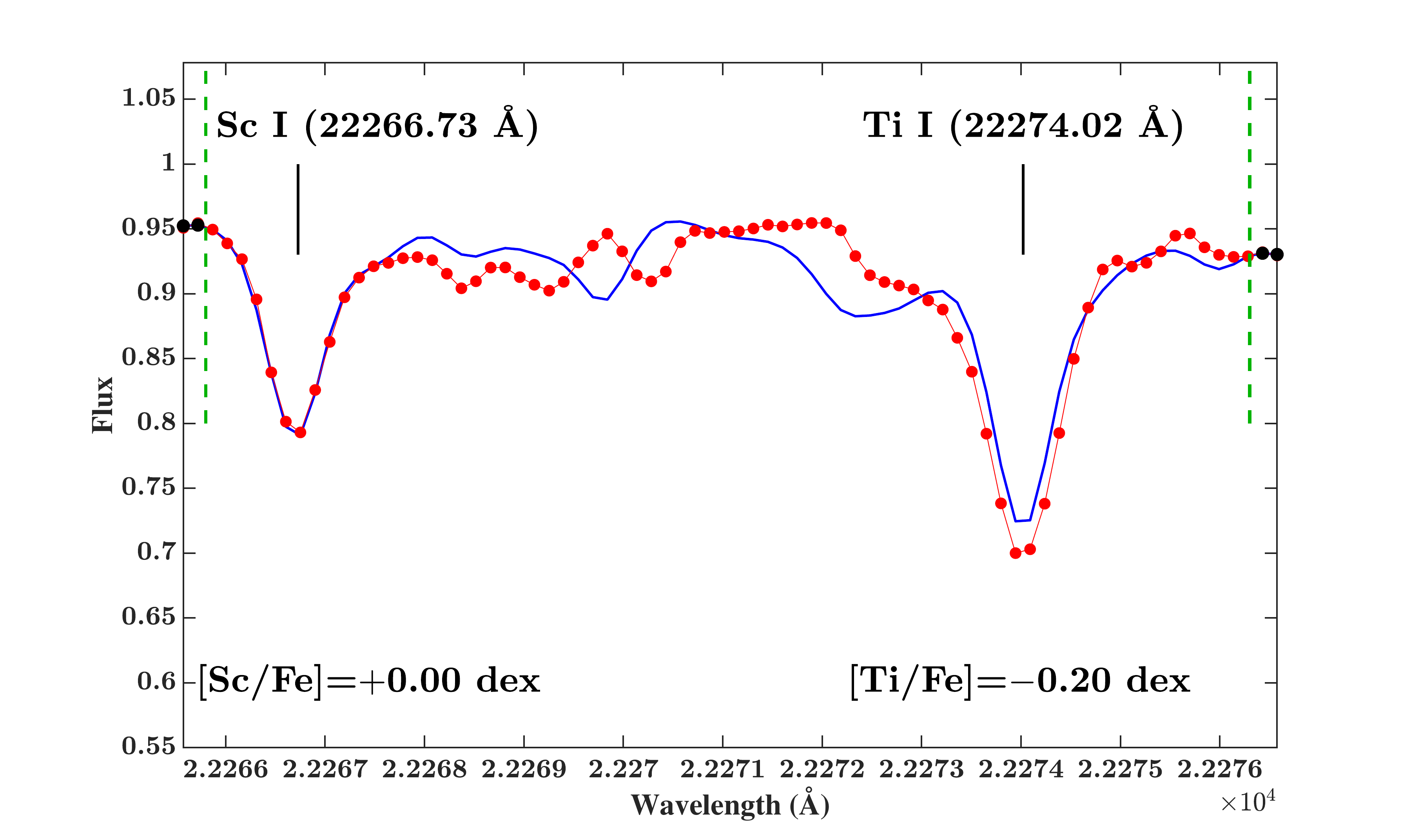}
\end{subfigure}
\vfill
\begin{subfigure}
  \centering
  \includegraphics[width=0.518\linewidth]{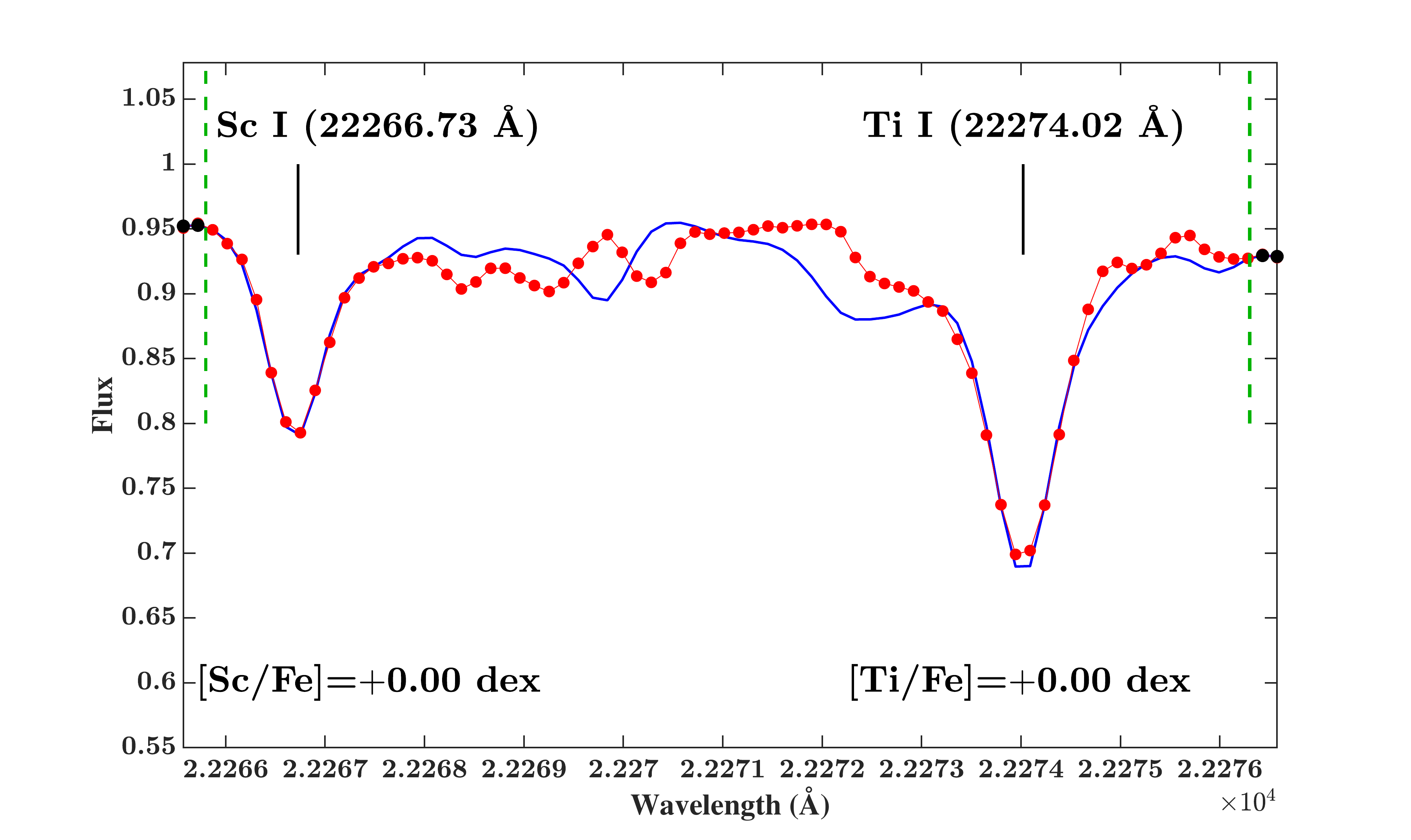}
\end{subfigure}
\hspace{-1.1cm}
\begin{subfigure}
  \centering
  \includegraphics[width=0.518\linewidth]{Abundace_Variation_Normalization_Ti_Sc_0.0-eps-converted-to.pdf}
\end{subfigure}
\vfill
\begin{subfigure}
  \centering
  \includegraphics[width=0.518\linewidth]{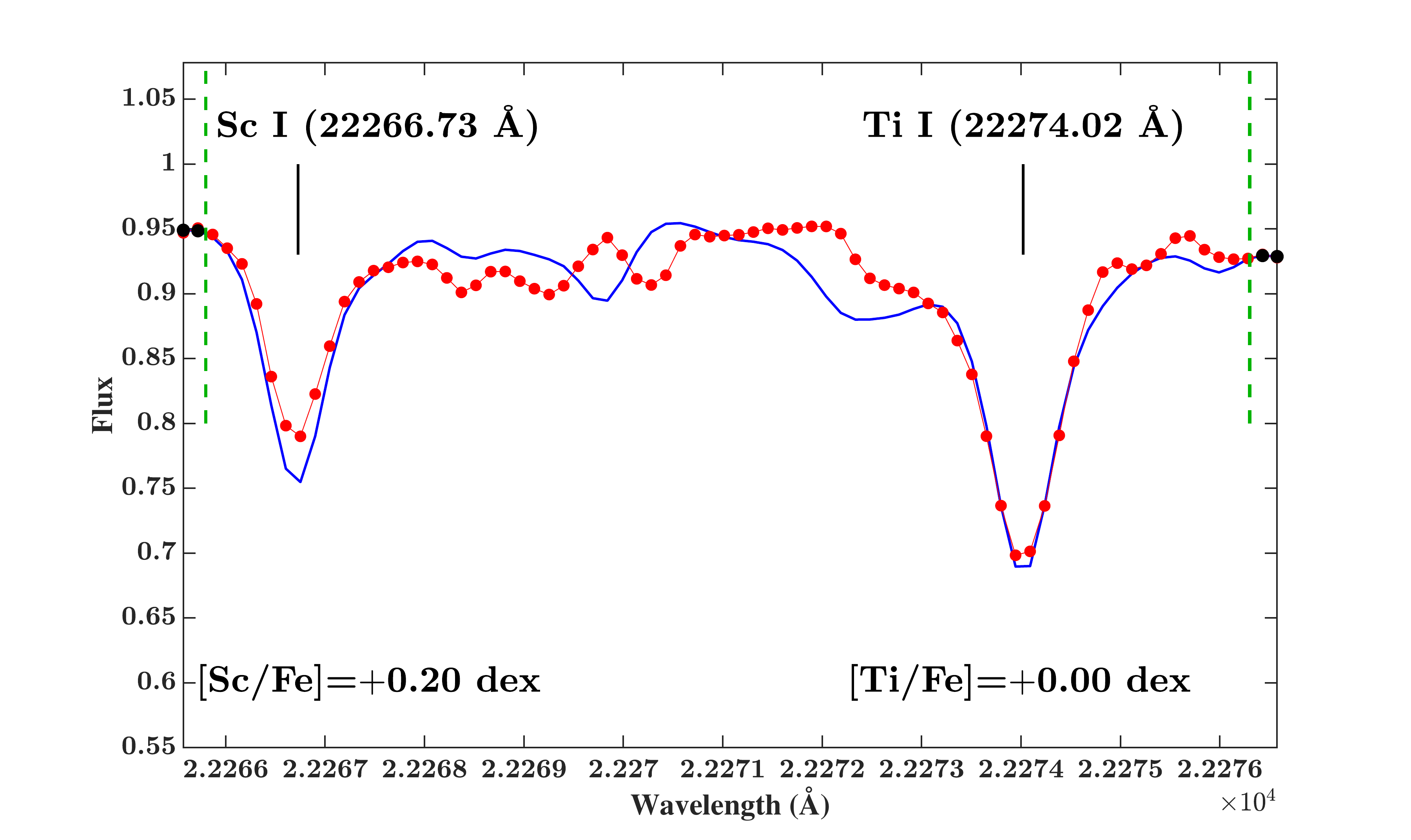}
\end{subfigure}
\hspace{-1.1cm}
\begin{subfigure}
  \centering
  \includegraphics[width=0.518\linewidth]{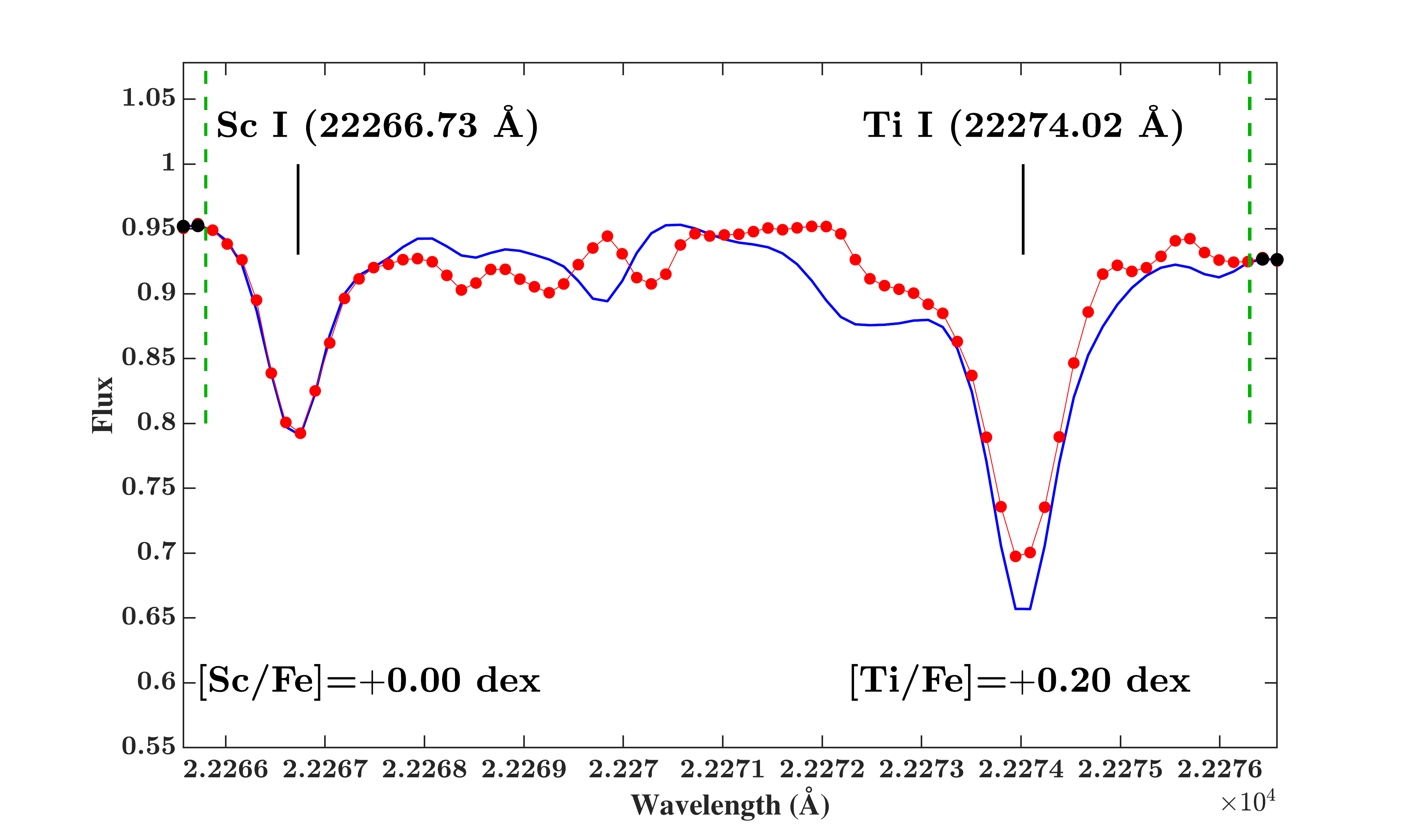}
\end{subfigure}
\caption{Comparison between the normalized observed spectrum (red lines and circles) of K2-18 and the model spectra (blue lines) associated with the target's parameters but varying abundances of the element Sc (left panels) and the element Ti (right panels), while assuming solar relative abundances for all other elements. The black circles (at the edges of the panels) show the normalizing points within the selected continuum/pseudocontinuum normalizing ranges that are separated from the inner spectral regions by green dashed lines.}
\label{fig:normalize_Ti_Sc}
\end{figure*}

\begin{deluxetable*}{lccl}\label{tab:line_data} 
 \tablenum{1}
\tablecaption{148 atomic and molecular lines selected for this analysis}  
\tablewidth{0pt}
\tabletypesize{\scriptsize}
\tablehead{
\colhead{Species} &  Central wavelength ({\AA}) & {\ensuremath{{\chi}^2}} window ({\AA})  & Comments}
\startdata
CO     & 23006.89 & 	23006.25-23007.40 &      \\
CO     & 23015.00 & 	23014.50-23015.50 &      \\
CO     & 23023.52 & 	23023.00-23024.10 &      \\
CO     & 23032.43 & 	23032.00-23033.15 &      \\
CO     & 23061.59 & 	23061.05-23062.10 &      \\
CO     & 23083.04 & 	23082.60-23083.50 &      \\
CO     & 23094.37 & 	23093.95-23094.80 &      \\
CO     & 23118.23 & 	23117.75-23118.75 &      \\
CO     & 23170.81 & 	23170.35-23171.40 &      \\
CO     & 23184.97 & 	23184.50-23185.45 &      \\
CO     & 23341.22 & 	23340.70-23341.95 &      \\
CO     & 23351.41 & 	23350.95-23352.05 &      \\
CO     & 23421.19 & 	23420.77-23421.70 &      \\
CO     & 23426.30 & 	23425.78-23426.70 &      \\
CO     & 23447.76 & 	23447.40-23448.25 &      \\
CO     & 23461.67 & 	23461.20-23462.10 &      \\
CO     & 23476.00 & 	23475.60-23476.40 &      \\
CO     & 23505.90 & 	23505.40-23506.55 &      \\
CO     & 23637.61 & 	23637.20-23638.00 &      \\
CO     & 23658.53 & 	23658.15-23658.95 &      \\
CO     & 23661.26 & 	23660.78-23661.73 &      \\
CO     & 23724.24 & 	23723.73-23724.75 &      \\
CO     & 23745.10 & 	23744.65-23745.60 &      \\
CO     & 23759.17 & 	23758.70-23759.70 &      \\
CO     & 24009.23 & 	24008.50-24009.75 &      \\
CO     & 24023.59 & 	24023.10-24024.00 &      \\
CO     & 24128.68 & 	24128.20-24129.15 &      \\
CO     & 24198.13 & 	24197.60-24198.70 &     \\
\hline
OH     & 15002.15 & 15001.85-15003.45 &   \\
OH     & 15003.12 & 15001.85-15003.45 &   \\
OH     & 15145.77 & 15145.50-15146.10 &   \\
OH     & 15147.94 & 15147.60-15148.30 &   \\
OH     & 15264.60 & 15264.30-15264.90 &   \\
OH     & 15266.17 & 15265.90-15266.45 &   \\
OH     & 15278.52 & 15278.16-15278.85 &   \\
OH     & 15281.05 & 15280.70-15281.41 &   \\
OH     & 15409.17 & 15408.90-15409.40 &   \\
OH     & 15419.46 & 15419.10-15419.72 &   \\
OH     & 15422.37 & 15421.97-15422.70 &   \\
OH     & 15558.02 & 15557.73-15558.37 &   \\
OH     & 15560.24 & 15559.90-15560.55 &   \\
OH     & 15568.78 & 15568.45-15569.11 &   \\
OH     & 15572.08 & 15571.72-15572.40 &   \\
OH     & 15626.70 & 15626.42-15627.70 &   \\
OH     & 15627.41 & 15626.42-15627.70 &   \\
OH     & 15651.90 & 15651.55-15652.20 &   \\
OH     & 15653.48 & 15653.20-15653.75 &   \\
OH     & 15719.70 & 15719.30-15720.10 &   \\
OH     & 15726.72 & 15726.44-15727.00 &   \\
OH     & 15755.52 & 15755.27-15755.77 &   \\
OH     & 15756.53 & 15756.20-15756.85 &   \\
OH     & 15884.90 & 15884.50-15885.30 &   \\
OH     & 15892.13 & 15891.80-15892.50 &   \\
OH     & 15893.53 & 15893.15-15893.80 &   \\
OH     & 15897.70 & 15897.30-15898.10 &   \\
OH     & 15910.42 & 15910.05-15910.80 &   \\
OH     & 15912.73 & 15912.33-15913.10 &   \\
\enddata
\end{deluxetable*}

\begin{deluxetable*}{lccl} 
 \tablenum{1}
\tablecaption{Continued}  
\tablewidth{0pt}
\tabletypesize{\scriptsize}
\tablehead{
\colhead{Species} &  Central wavelength ({\AA}) & {\ensuremath{{\chi}^2}} window ({\AA}) &  Comments}
\startdata
OH     & 16036.89 & 16036.43-16037.20 &   \\
OH     & 16038.54 & 16038.20-16038.85 &   \\
OH     & 16052.76 & 16052.43-16053.10 &   \\
OH     & 16055.46 & 16055.10-16055.78 &   \\
OH     & 16065.05 & 16064.80-16065.40 &   \\	
OH     & 16069.52 & 16069.17-16069.90 &   \\
OH     & 16190.13 & 16189.80-16190.50 &   \\
OH     & 16192.13 & 16191.80-16192.40 &   \\
OH     & 16207.19 & 16206.70-16207.50 &   \\
OH     & 16247.88 & 16247.53-16248.27 &   \\
OH     & 16260.15 & 16259.74-16260.56 &   \\
OH     & 16346.18 & 16345.81-16346.57 &   \\
OH     & 16352.22 & 16351.75-16352.65 &   \\
OH     & 16354.58 & 16354.22-16354.96 &   \\
OH     & 16364.59 & 16364.20-16364.95 &   \\
OH     & 16368.13 & 16367.78-16368.53 &   \\
OH     & 16448.05 & 16447.70-16448.50 &   \\ 
OH     & 16450.37 & 16449.98-16450.80 &   \\
OH     & 16456.04 & 16455.70-16456.40 &   \\	
OH     & 16471.15 & 16470.82-16471.50 &   \\
OH     & 16523.50 & 16523.15-16523.80 &   \\	
OH     & 16526.25 & 16525.90-16526.60 &   \\
OH     & 16534.58 & 16534.28-16534.93 &   \\
OH     & 16538.59 & 16538.10-16538.88 &   \\
OH     & 16581.27 & 16580.95-16581.70 &   \\ 
OH     & 16582.32 & 16581.98-16582.60 &   \\ 
OH     & 16649.95 & 16649.60-16650.40 &   \\ 
OH     & 16654.65 & 16654.32-16654.98 &   \\ 
OH     & 16655.99 & 16655.65-16656.37 &   \\ 
OH     & 16662.20 & 16661.87-16662.55 &   \\
OH     & 16704.36 & 16703.95-16704.90 &   \\
OH     & 16866.69 & 16866.30-16867.05 &   \\
OH     & 16879.09 & 16878.70-16879.52 &   \\
OH     & 16895.18 & 16894.68-16895.64 &   \\   
OH     & 16902.73 & 16902.32-16903.17 &   \\
OH     & 16904.28 & 16903.90-16904.75 &   \\
OH     & 16905.63 & 16905.25-16905.95 &   \\
OH     & 16909.29 & 16908.90-16909.75 &   \\
OH     & 17052.20 & 17051.85-17052.60 &   \\
OH     & 17066.13 & 17065.77-17066.50 &   \\
OH     & 17069.48 & 17069.15-17069.78 &   \\
OH     & 17094.52 & 17094.20-17094.95 &   \\
OH     & 17096.39 & 17095.97-17096.80 &   \\
OH     & 17239.72 & 17239.45-17240.00 &   \\ 
OH     & 17767.06 & 17766.75-17767.35 &   \\
\hline
Na I     & 22083.66 &   22082.35-22085.00 &  The combination of four Na I lines: \\
{} & {} & {} &  22083.617, 22083.627*, 22083.684*, 22083.694*, \\
{} & {} & {} &  including three HFS lines \\
\hline
Mg I     & 15040.25 & 15039.80-15040.65 &          \\
Mg I     & 15047.71 & 15047.20-15048.10 &          \\
Mg I     & 15765.84 & 15765.30-15766.32 &   Blended with two Mg I lines:     \\
{} & {} & {} & 15765.645, 15765.747, \\
{} & {} & {} &   significantly weaker than the main, central line (i.e., 15765.84),     \\
{} & {} & {} &  the three blended lines have different J values of  the upper levels \\
Mg I     & 17108.63 & 17108.10-17109.05 &         \\
\hline
Al I     & 16718.96 & 16718.10-16719.70 &  The combination of six Al I lines:   \\
{} & {} & {} &  16718.911, 16718.925*,  16718.943*, \\
{} & {} & {} &  16718.945*, 16718.963*, 16718.990*, \\ 
{} & {} & {} &  including five HFS lines\\
Al I     & 16750.56	& 16750.00-16751.10 &   The combination of 12 Al I lines: \\
{} & {} & {} &  16750.455, 16750.539*, 16750.550*, 16750.608*,  \\
{} & {} & {} &   16750.616*, 16750.627*, 16750.660*, 16750.665*, \\
{} & {} & {} &    16750.673*, 16750.698*, 16750.703*, 16750.717* \\
{} & {} & {} &  including 11 HFS lines \\
\enddata
\tablecomments{* denotes a line resulted from  hyperfine structure (HFS) splitting.}
\end{deluxetable*}

\begin{deluxetable*}{lccl} 
 \tablenum{1}
\tablecaption{Continued}  
\tablewidth{0pt}
\tabletypesize{\scriptsize}
\tablehead{
\colhead{Species} &  Central wavelength ({\AA}) & {\ensuremath{{\chi}^2}} window ({\AA})  & Comments}
\startdata
K I      & 15168.38 & 15167.95-15168.80 &           \\
\hline
Ca I     & 19853.09 & 19852.57-19853.70 &         \\
Ca I     & 19933.73 & 19933.20-19934.30 &          \\
Ca I     & 22607.94 & 22607.20-22608.65 &           \\
\hline
Sc I     & 22266.73 & 22266.25-22267.25 &  The combination of six Sc I lines:  \\
{} & {} & {}  &  22266.533,  22266.637*, 22266.715*, \\
{} & {} & {}  & 22266.739*,  22266.871*, 22266.975*, \\
{} & {} & {}  &  including five HFS lines    \\
\hline
Ti I     & 15334.85 &	15334.47-15335.20 & Blended with three  weak Ti I lines:   \\
{} & {} & {}  & 15334.139, 15335.039, 15335.458, \\
{} & {} & {}  & too weak to influence the shape of the main, central line (i.e., 15334.85), \\
{} & {} & {}  & the four blended lines have different J values of the lower and/or upper levels  \\ 
Ti I     & 15715.57 &	15715.10-15716.20 &  Blended with four weak Ti I:    \\
{} & {} & {} &   15715.758, 15715.887, 15716.008, 15716.484, \\
{} & {} & {} &   too weak to influence the shape of the main, central line (i.e., 15715.57),  \\
{} & {} & {}  &  the five blended lines have different J values of the lower and/or upper levels  \\ 
Ti I     & 21782.94 &	21782.20-21783.75 &  Blended with three Ti I lines:      \\
{} & {} & {}  &  21782.555, 21782.560, 21782.996,   \\
{} & {} & {}  &  too weak to influence the shape of the main, central line (i.e., 21782.94) \\
{} & {} & {}  &  the four blended lines have different J values of the lower and/or upper levels \\ 
Ti I     & 21897.39 &	21896.75-21898.15 &       \\
Ti I     & 22004.51 &	22004.00-22004.95 &       \\ 
Ti I     & 22211.24 &	22210.55-22211.95 &  Blended with one  Ti I line:    \\
{} & {} & {} & 22210.631, \\
{} & {} & {}  & too weak to influence the shape of the main, central line (i.e., 22211.24), \\
{} & {} & {}  & the two blended lines have different J value  of the lower levels \\
Ti I     & 22232.86 &	22232.20-22233.50 &       \\
Ti I     & 22274.02 &	22273.45-22274.55 &       \\
Ti I     & 22963.33 &   22962.67-22963.94 &       \\
Ti I     & 23441.48 &	23441.15-23441.95 &   Blended with two  Ti I:   \\
{} & {} & {}  &  23440.630,  23441.669,  \\
{} & {} & {}  & too weak to influence the shape of the main line, central (i.e., 23441.48), \\
{} & {} & {}  & the three blended lines have different J values of the lower and/or upper levels \\ 
\hline
FeH    & 15872.67 &	15872.31-15873.00 &      \\
FeH    & 15915.94 &	15915.70-15916.22 &      \\
FeH    & 15945.71 &	15945.39-15946.00 &      \\
FeH    & 15993.22 &	15992.93-15993.60 &      \\
FeH    & 16058.56 &	16058.27-16058.89 &      \\
FeH    & 16067.85 &	16067.60-16068.20 &      \\
FeH    & 16172.62 &	16172.35-16173.00 &      \\
FeH    & 16182.95 &	16182.70-16183.25 &      \\
FeH    & 16184.38 &	16184.10-16184.80 &      \\
FeH    & 16249.70 &	16249.30-16249.98 &      \\
FeH    & 16319.36 &	16319.08-16319.70 &      \\
FeH    & 16330.67 &	16330.20-16330.93 &      \\
FeH    & 16361.74 &	16361.45-16362.08 &      \\
FeH    & 16466.93 &	16466.45-16467.20 &      \\
FeH    & 16682.00 &	16681.70-16682.30 &      \\
FeH    & 16735.42 &	16735.15-16735.65 &      \\
FeH    & 16738.29 &	16737.97-16738.58 &      \\
FeH    & 16796.38 &	16796.05-16796.68 &      \\
FeH    & 16862.14 &	16861.77-16862.42 &      \\
FeH    & 16922.75 &	16922.40-16923.00 &      \\
FeH    & 17068.40 &	17068.05-17068.75 &      \\
FeH    & 17277.76 &	17277.40-17278.10 &      \\
FeH    & 17293.38 &	17292.90-17293.70 &      \\
FeH    & 17544.47 &	17544.12-17544.75 &      \\
\enddata
\tablecomments{* denotes a line resulted from  hyperfine structure (HFS) splitting.}
\end{deluxetable*}

 \begin{deluxetable*}{lc|c|cc|cc|cc|cc|c|c|c}
 \tablenum{2}
 \label{tab:results} 
\tablecaption{The chemical abundances and their corresponding uncertainties for the ten studied elements}  
\tablewidth{0pt}
\tabletypesize{\scriptsize}
\tablehead{
\colhead{Species} &  \rm{N} & {[X/H]} &
\multicolumn{2}{c|}{${\Delta}T_{\rm eff}$} & \multicolumn{2}{c|}{$\Delta$[M/H]} & \multicolumn{2}{c|}{$\Delta$log($g$)} & \multicolumn{2}{c|}{$\Delta\xi$} &  $\sigma_{\rm{sys}}$  &  $\sigma_{\rm{ran}}=\rm{std}/\sqrt{N}$     & {${\sigma}{\rm [X/H]}_{\rm tot}$}\\
\colhead{} & {}
{} & 
{} & 
$-$85 & +85 & {$-$0.10} & {+0.10} & {$-$0.10} & {+0.10} & {$-$0.10} &{+0.10} & {} & { } \\
\colhead{} & {} &  {(dex)} &  {(K)} & {(K)} & {(dex)} & {(dex)} & {(dex)} & {(dex)} & {km/s} & {km/s} & {(dex)} & {(dex)} & {(dex)} 
}
\startdata
C (CO) & 28   &	+0.104  & $-$0.003   & $-$0.004   &	+0.004   &	$-$0.006   &	$-$0.084   & +0.081   & +0.027 &  $-$0.029  &  0.088 &   0.011   & 0.089\\
O (OH) & 74   &	+0.080	& +0.014     &	$-$0.019  &	$-$0.005 & +0.007      &  $-$0.079     & +0.077	  & +0.018 &  $-$0.021  &  0.083 &   0.002   & 0.083\\
Na & 1        &	+0.066	& +0.064     &	$-$0.076  &	+0.073   &	$-$0.085   &	$-$0.080   & +0.076   & +0.010 &  $-$0.012	&  0.132 &    --     & 0.132\\
Mg	& 4       & +0.043	& +0.174     &	$-$0.142  &	+0.005   &	+0.023	   & 	$-$0.075   & +0.096   & +0.005 &  $-$0.004  &  0.181 &    --     & 0.181\\
Al	& 2       & +0.105  & +0.177	 & $-$0.156   & +0.056	 & $-$0.038    &	$-$0.130   & +0.133   & +0.003 &  $-$0.007  &  0.218 &    --     & 0.218\\
K & 1         & +0.040	& $-$0.019   &	+0.025	  & +0.002   &	$-$0.007   &	$-$0.026   & +0.025	  & +0.002 &  $-$0.003  &  0.035 &    --     & 0.035\\
Ca & 3	      & +0.074  & +0.183     &	$-$0.176  & +0.018	 & $-$0.007    &    $-$0.138   & +0.122   & +0.012 &  $-$0.002	&  0.223 &    --     & 0.223\\
Sc & 1        & +0.134  & +0.039	 & $-$0.028   &	$-$0.002 &	+0.001     &	$-$0.083   & +0.083   &	+0.003 &  $-$0.006  &  0.090 &    --     & 0.090\\
Ti & 10       & +0.088  & +0.105     &	$-$0.091  &	+0.025	 & $-$0.028    &    $-$0.103  & +0.103    &	+0.011 &  $-$0.016  &  0.145 &   0.016    & 0.146\\
Fe (FeH) & 24 & $-$0.033 & +0.051    & $-$0.048   &	+0.053   & +0.007      &	$-$0.082   & +0.100   &	+0.012 &  $-$0.023	&  0.110 &   0.012   & 0.111\\
\enddata
\end{deluxetable*}

\begin{figure*}[hbt!]
    \centering
    \includegraphics[width=1\linewidth]{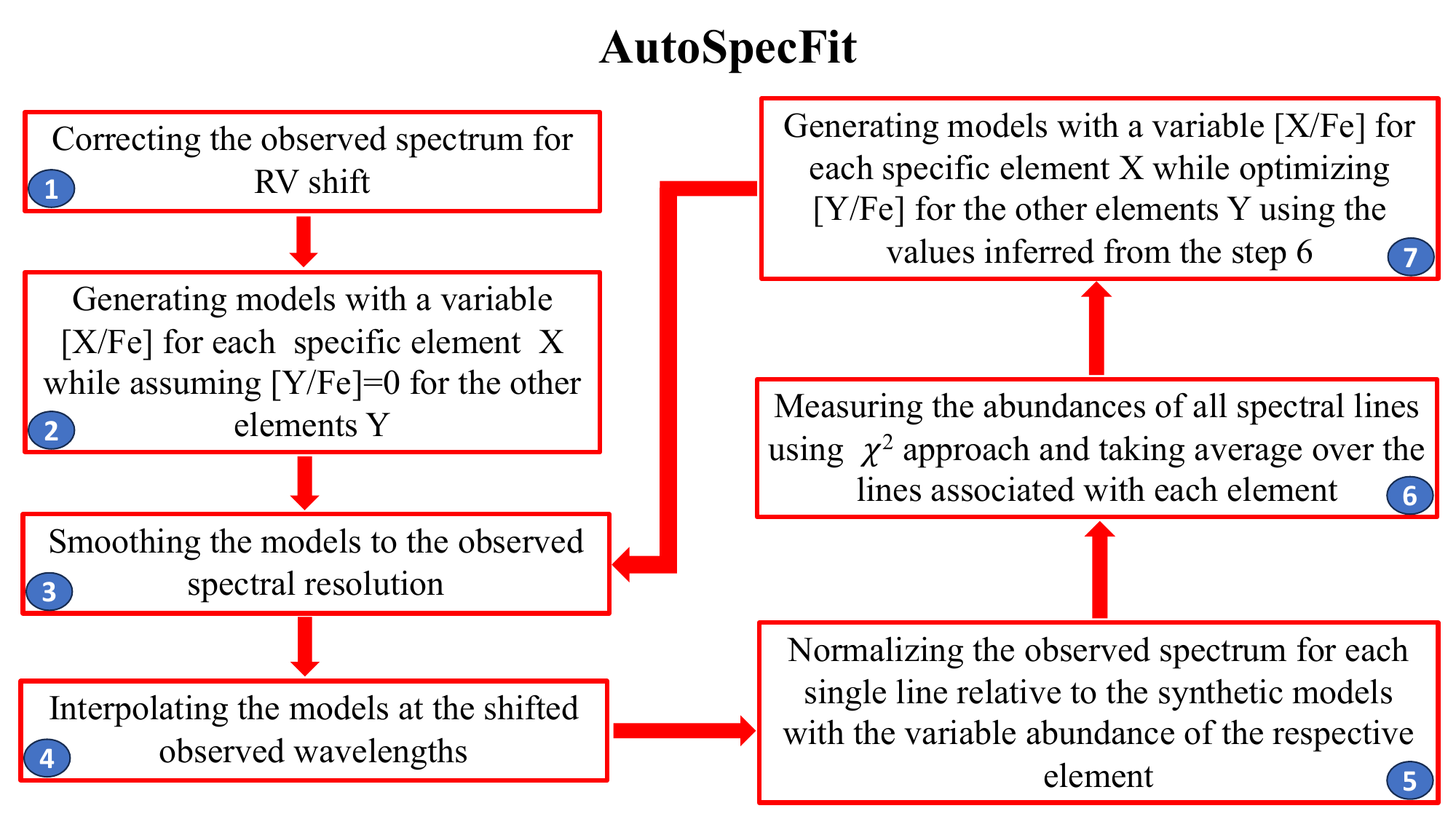}
    \caption{The flowchart of the AutoSpecFit performance from step 1 to step 7. The first two steps are run only in the first iteration. The pipeline returns back to step 3 to start  the next iteration.}
\label{fig:flowchart}
\end{figure*}

\begin{figure*}[hbt!]
    \centering
    \includegraphics[width=1\linewidth]{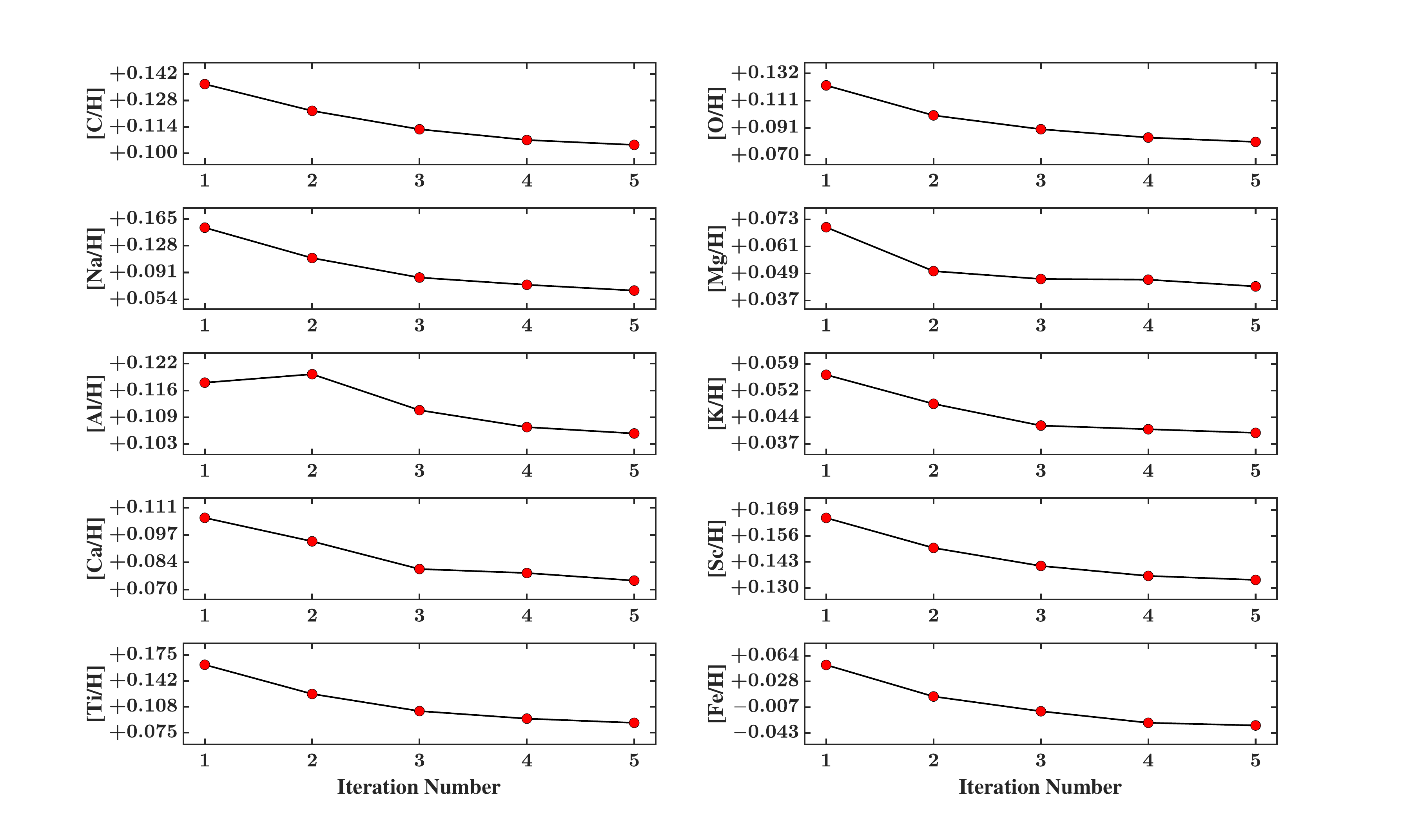}
    \caption{The abundance of the 10 analyzed elements as a function of the iteration number. The abundances are inferred using the models associated with the target's parameters, i.e., $T_{\rm eff}$ = 3547 K, [M/H] = 0.17 dex, log($g$) = 4.90 dex, and $\xi$ = 1.0 km/s. The total number of iterations is 5.}
\label{fig:iteration}
\end{figure*}

%---------------------------

\begin{figure*}[hbt!]
\centering
\begin{subfigure}
  \centering
  \includegraphics[width=0.40\linewidth]{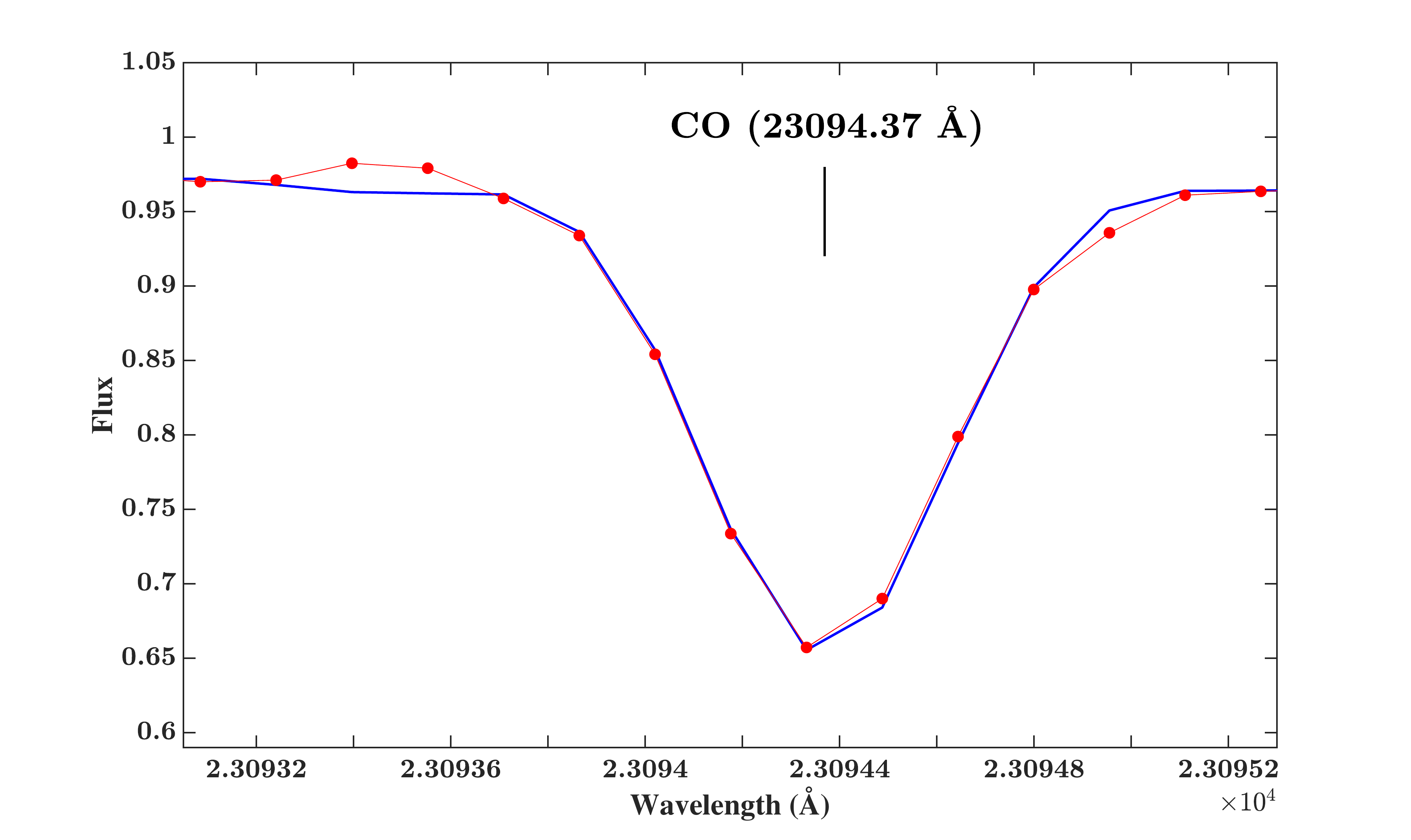}
\end{subfigure}
\begin{subfigure}
  \centering
  \includegraphics[width=0.40\linewidth]{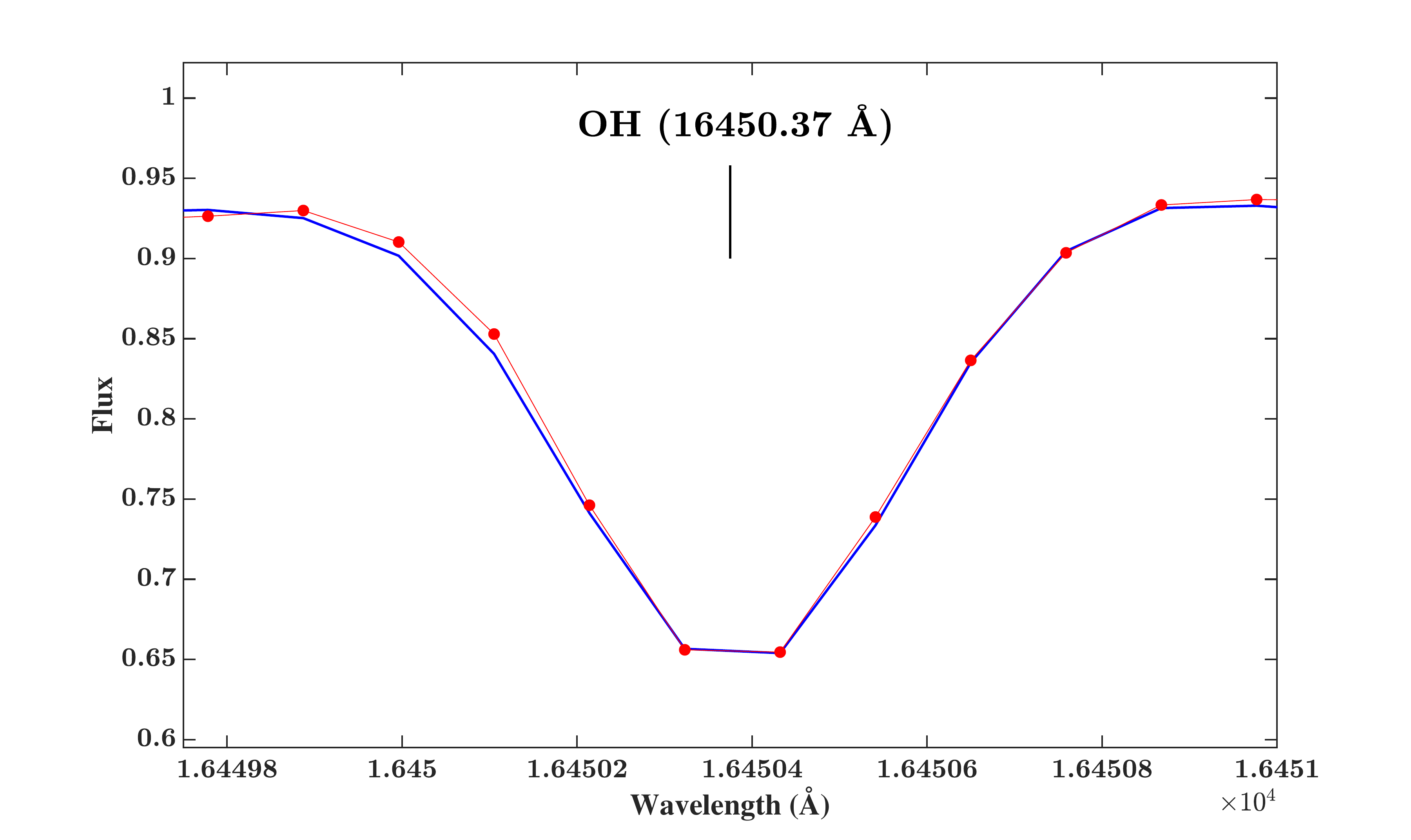}
\end{subfigure}
\vfill
\begin{subfigure}
  \centering
  \includegraphics[width=0.40\linewidth]{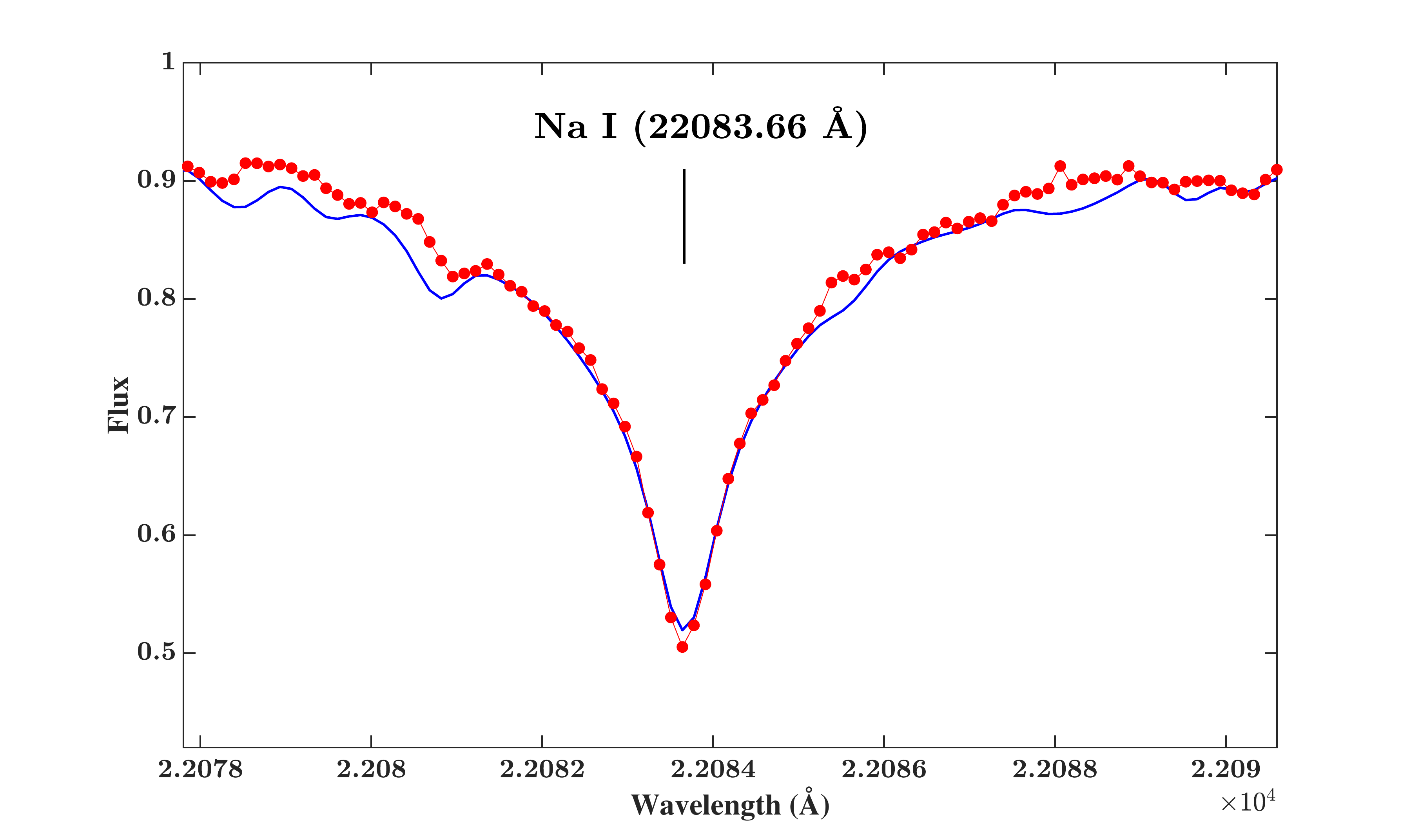}
\end{subfigure}
\begin{subfigure}
  \centering
  \includegraphics[width=0.40\linewidth]{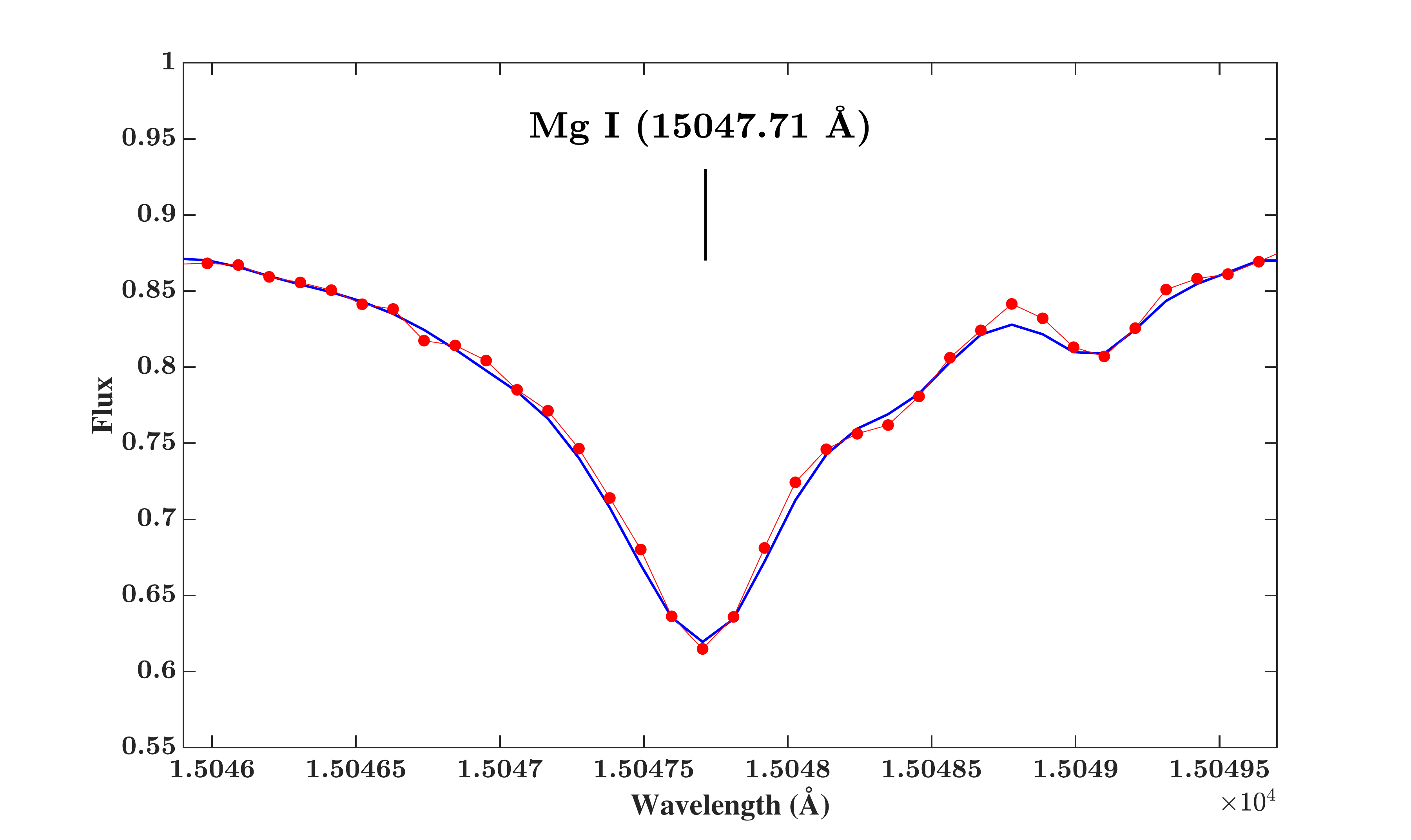}
\end{subfigure}
\vfill
\begin{subfigure}
  \centering
  \includegraphics[width=0.40\linewidth]{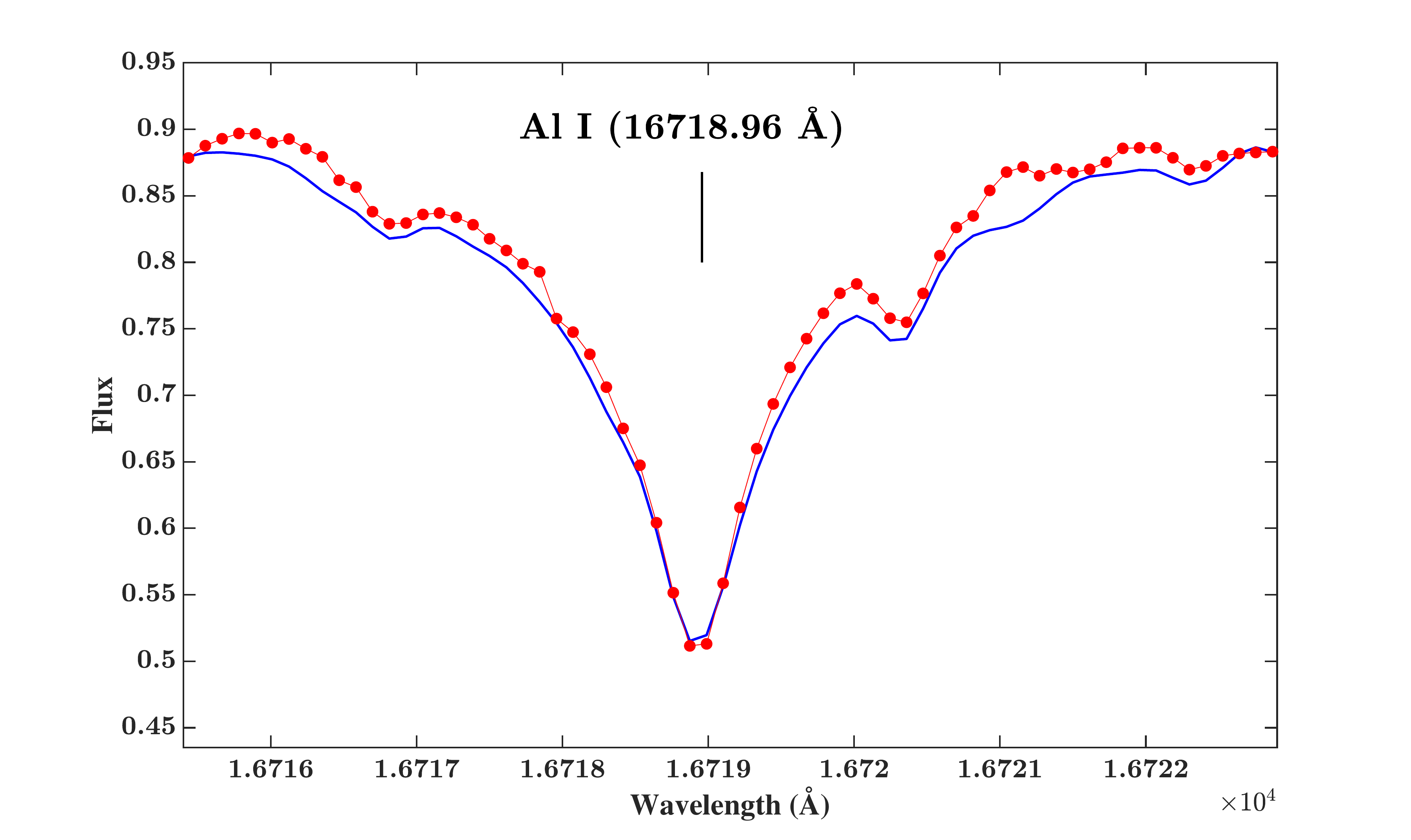}
\end{subfigure}
\begin{subfigure}
  \centering
  \includegraphics[width=0.40\linewidth]{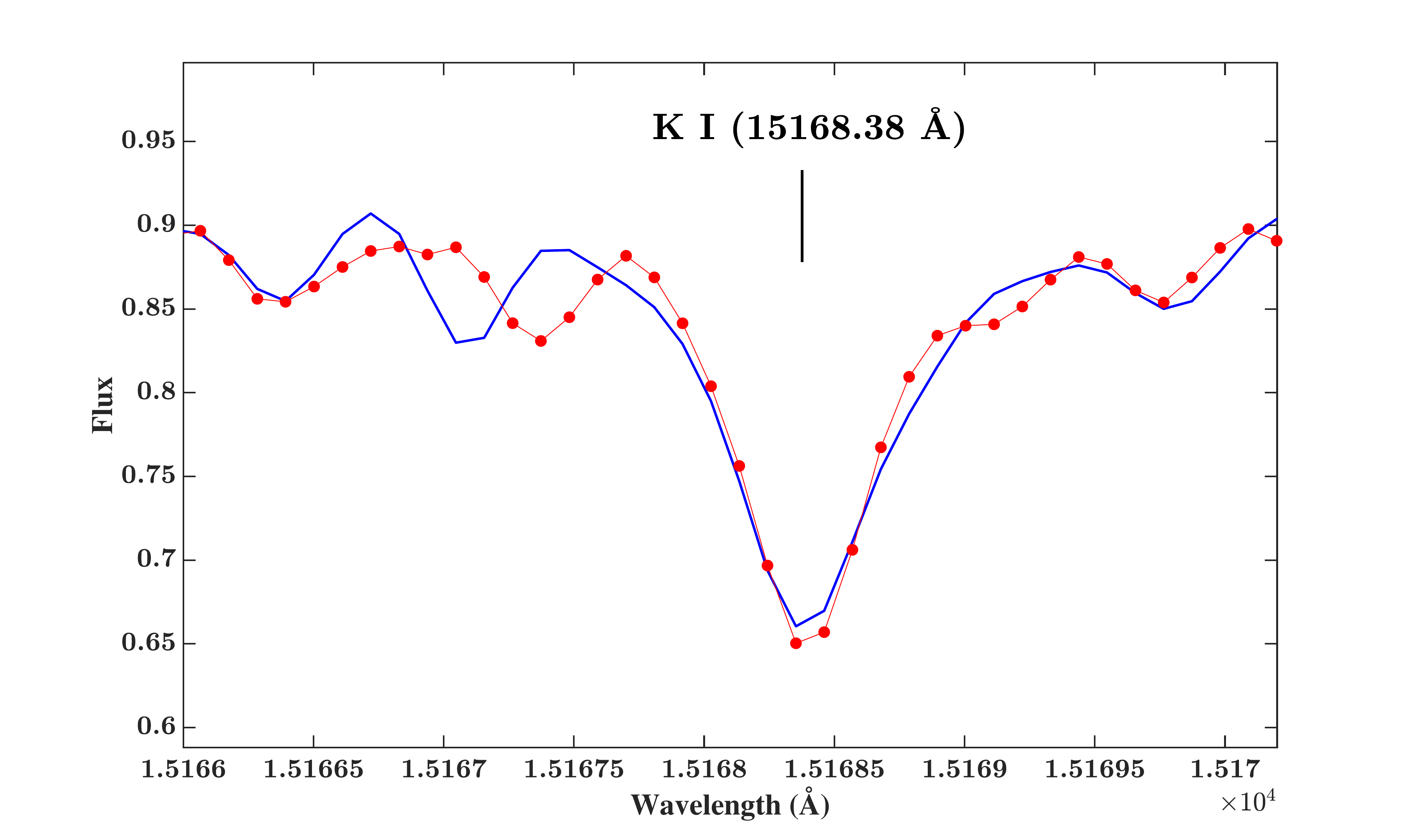}
\end{subfigure}
\vfill
\begin{subfigure}
  \centering
  \includegraphics[width=0.40\linewidth]{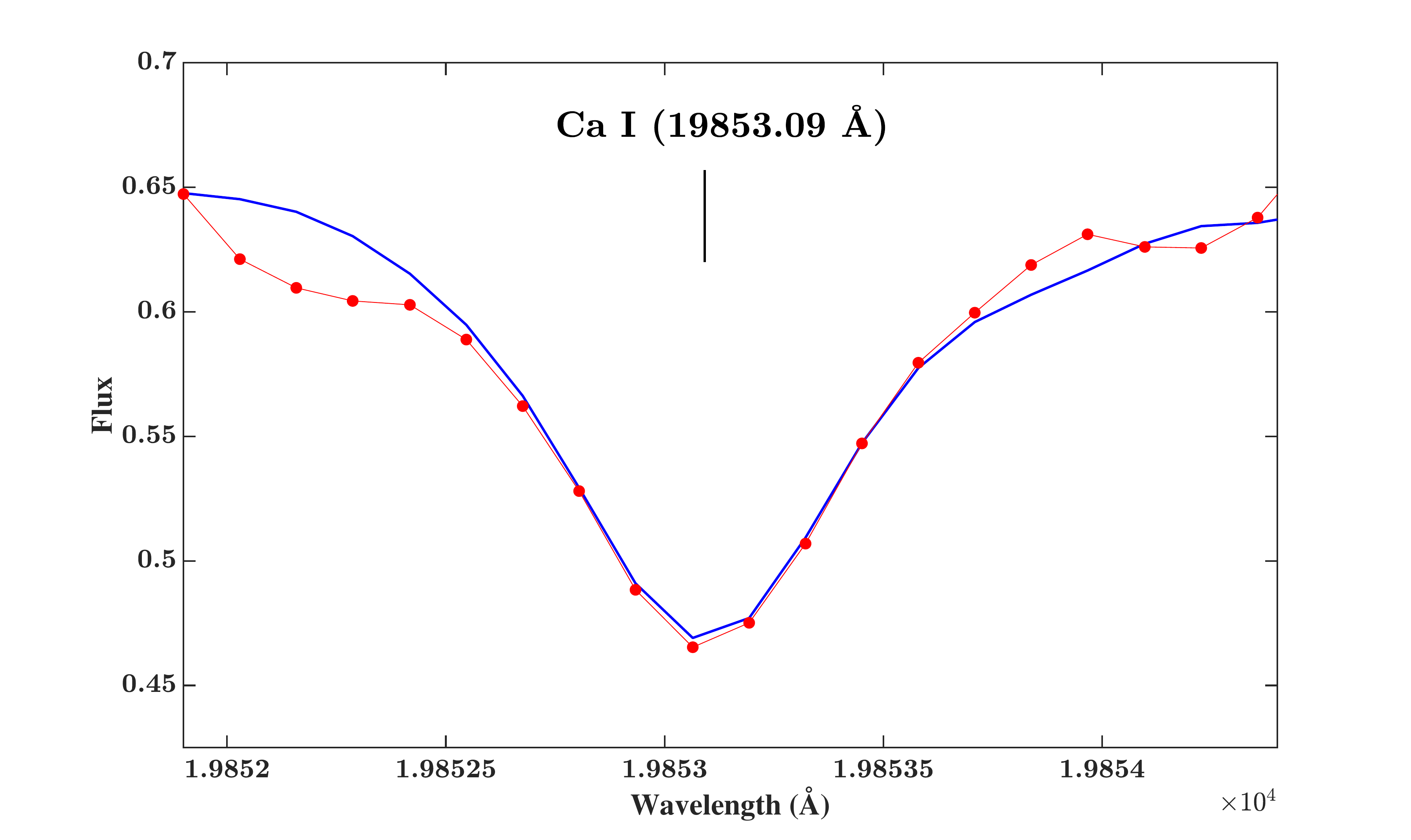}
\end{subfigure}
\begin{subfigure}
  \centering
  \includegraphics[width=0.40\linewidth]{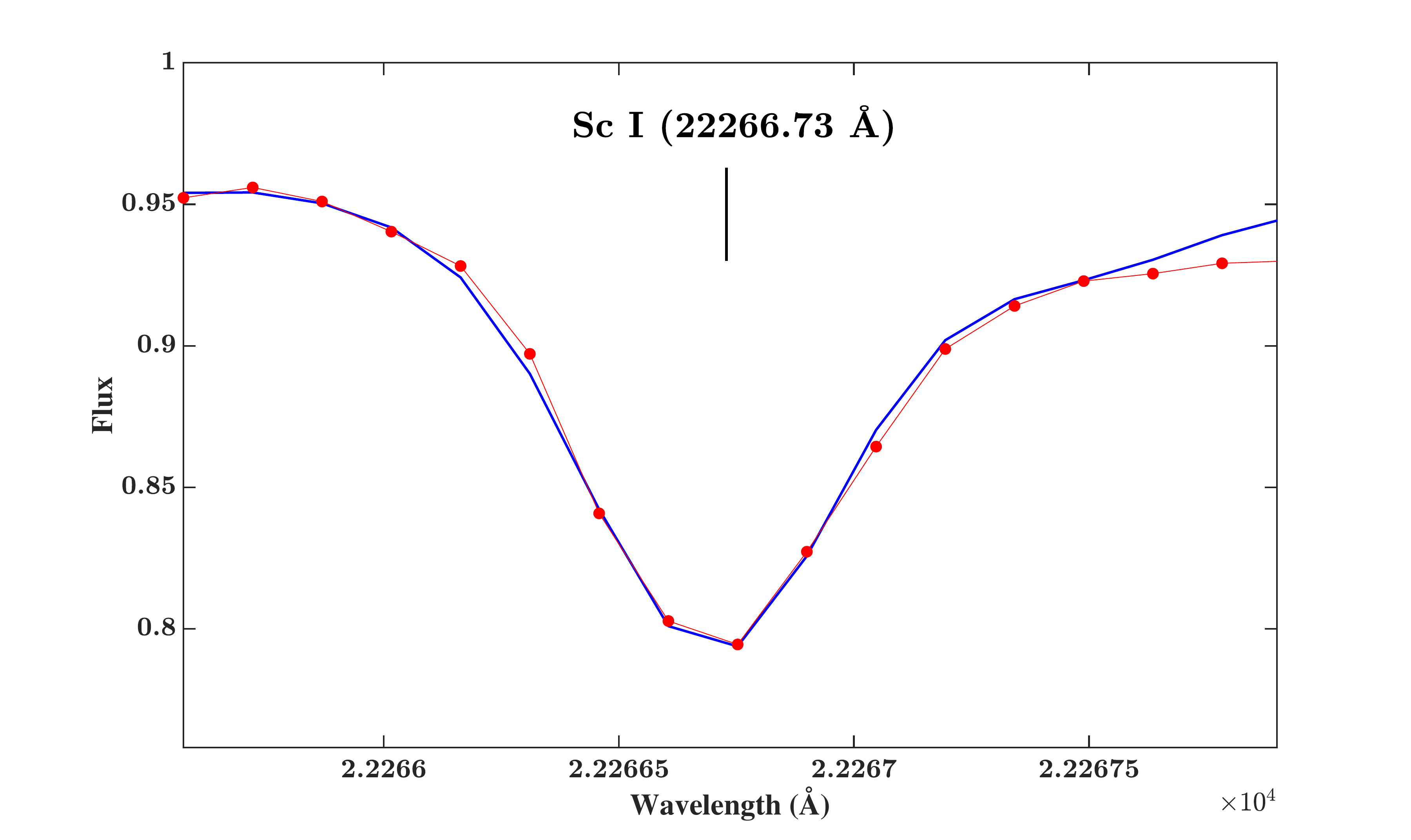}
\end{subfigure}
\vfill
\begin{subfigure}
  \centering
  \includegraphics[width=0.40\linewidth]{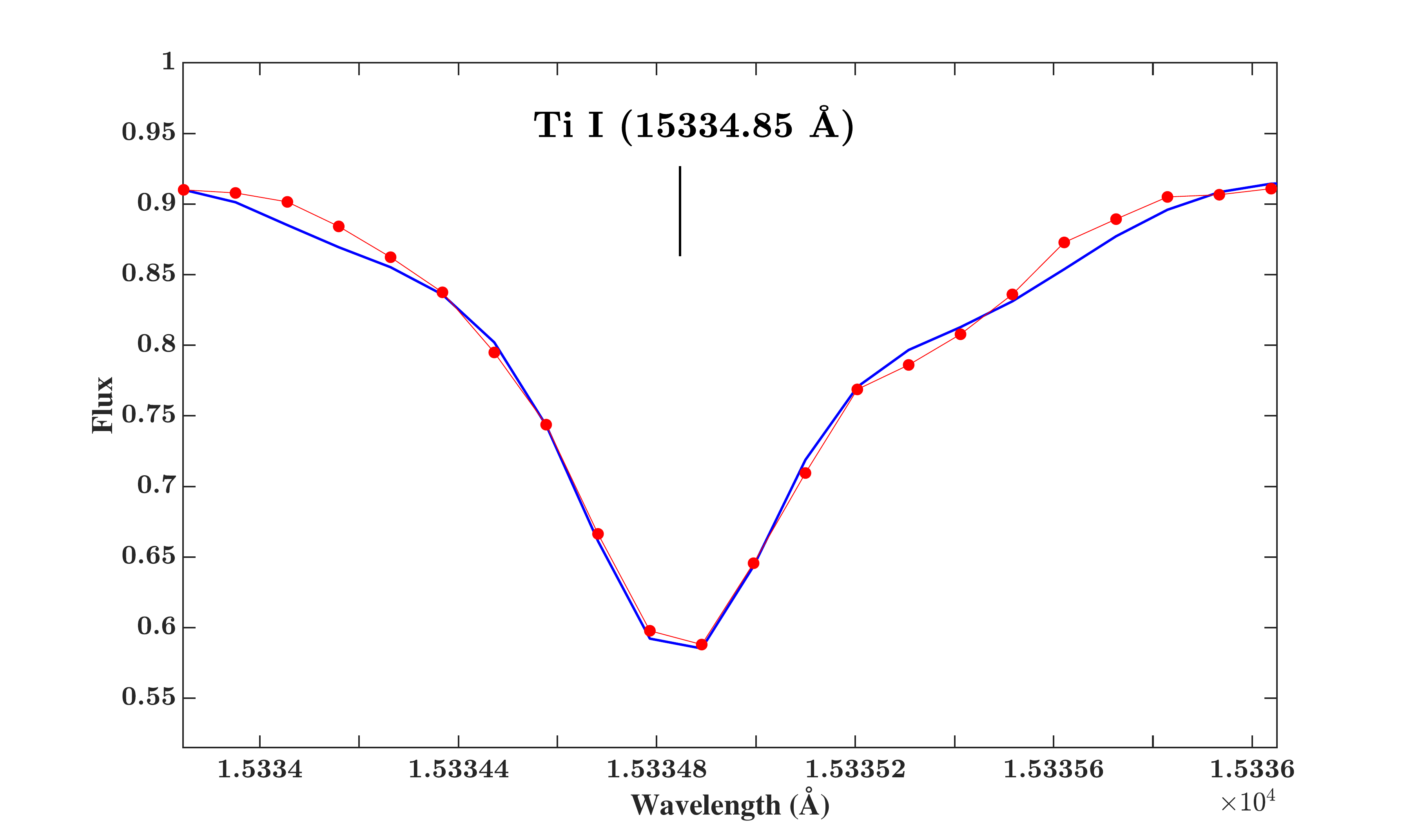}
\end{subfigure}
\begin{subfigure}
  \centering
  \includegraphics[width=0.40\linewidth]{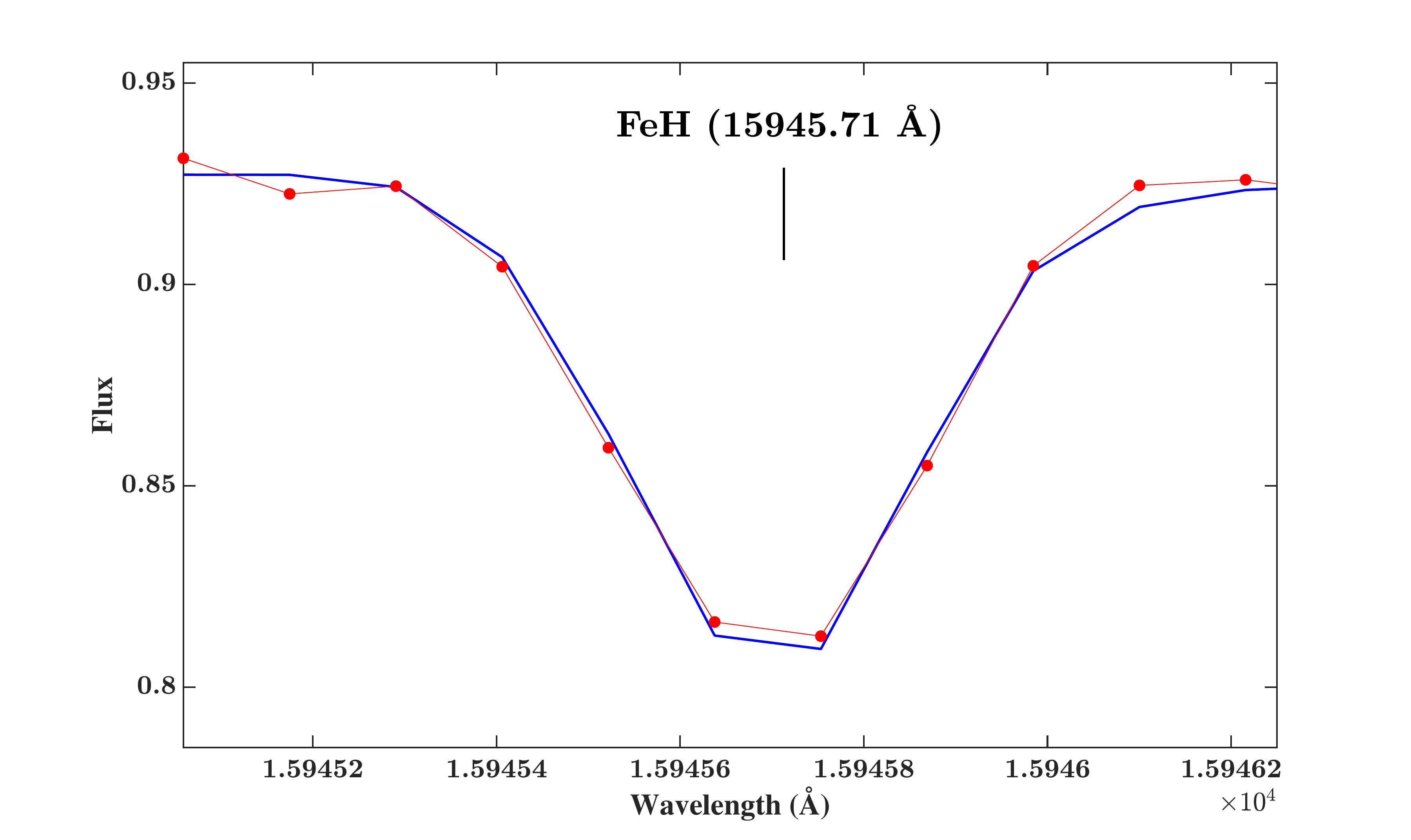}
\end{subfigure}
\caption{Comparison between the normalized observed spectrum (red lines and circles) and the final best-fit model (blue lines) over 10 spectral lines corresponding to the 10 analyzed elements.}
\label{fig:best_fit}
\end{figure*}

\begin{figure*}[hbt!]
    \centering
    \includegraphics[width=0.9\linewidth]{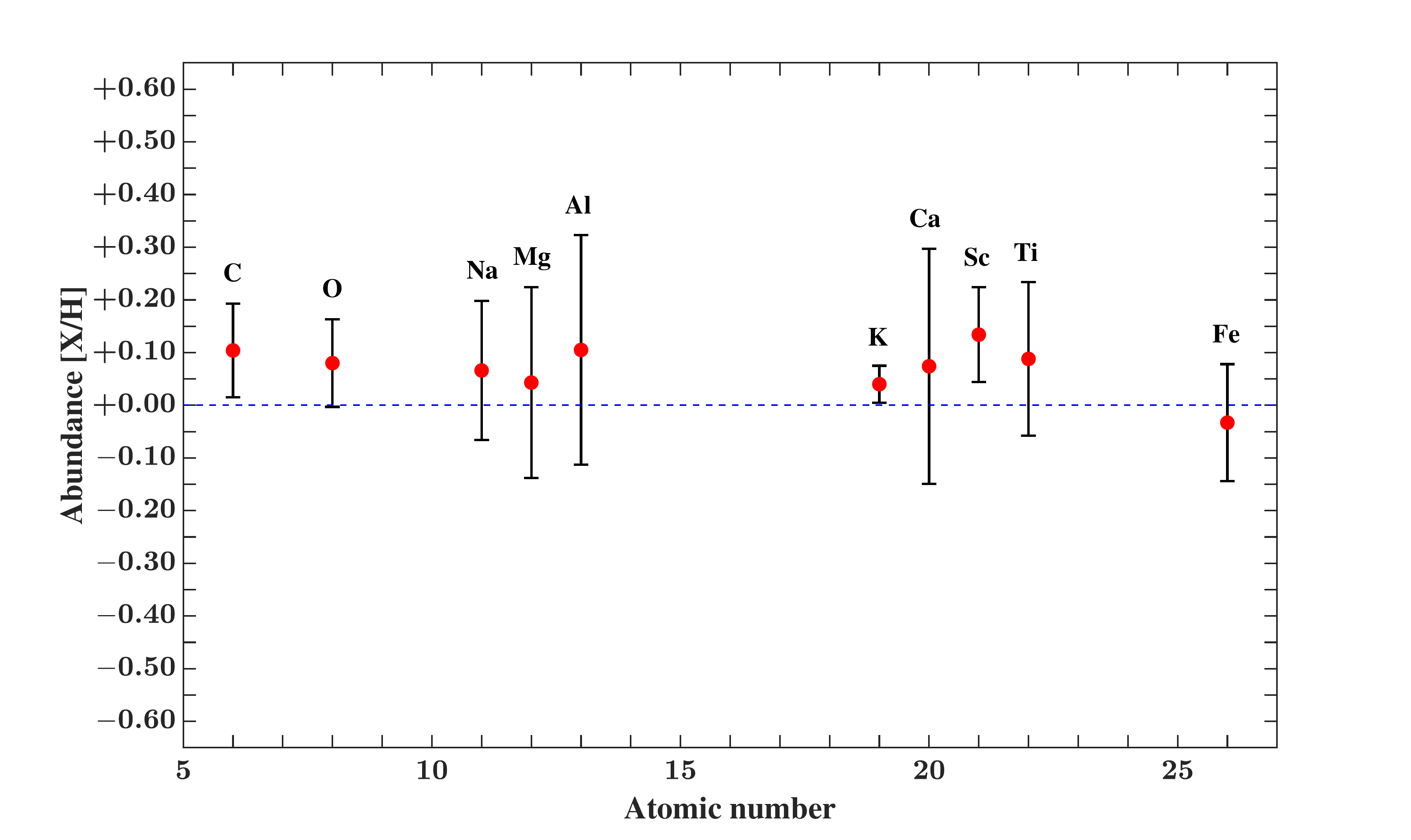}
    \caption{The final inferred  abundances of the 10 analyzed elements versus their corresponding atomic number. The error bars show the uncertainty of the abundances (as presented in the last column of Table 2). The blue dashed line shows the zero abundance level.}
\label{fig:abundance_atomicnumber}
\end{figure*}

\section{Summary and Conclusion}  \label{sec:summary}

\subsection{High-resolution Spectroscopic Analysis of K2-18}
We introduce AutoSpecFit, a new automatic line-by-line synthetic model fitting code, to measure the chemical abundances of cool dwarfs. The code performs a series of iterative {\ensuremath{{\chi}^2}} minimization processes and allows Turbospectrum to generate the synthetic spectra required for each iteration, which are optimized using the abundances inferred from the previous iteration. We illustrate how the abundances of different elements are dependent on each other and pass through multiple iterations to reach their final abundances that are globally consistent. Our abundance analysis offers a technique that carefully takes into account the complex dependency between different elements when varying their abundances in a timely manner. In addition, we present our method for continuum/pseudocontinuum normalization to make a meaningful comparison between the observed and model spectrum in the  {\ensuremath{{\chi}^2}} minimization. Since the continuum level cannot be identified in many spectral regions of cool dwarfs, we normalize the observed spectrum relative to synthetic, continuum-normalized spectra using several wavelength data points around the spectral lines of interest. 

We apply our technique to the high-resolution IGRINS H- and K-band spectra of the sub-Neptune K2-18's host M dwarf and measure the abundances of 10 elements, C (using CO lines), O (using OH lines), Na, Mg, Al, K, Ca, Sc, Ti, and Fe (using FeH lines), along with their detailed error analysis. We find near-solar abundances and carbon-to-oxygen ratio, C/O=0.568 $\pm$ 0.026. We also obtain the abundance ratios of some key planet-building elements, such as Al/Mg, Ca/Mg, and Fe/Mg. We emphasize that the accuracy of inferred abundances depends on the accuracy of  the input physical parameters as well as the normalization procedure. In particular, more accurate parameters, especially effective temperature, would lead to more accurate elemental abundances.

\subsection{Star-Planet Connection}

The exoplanet K2-18~b has been targeted by several JWST programs, and its atmosphere is  being characterized  more accurately than from previous studies, for example, using HST observations.  Historically, exoplanet abundances have been derived assuming Solar abundances, however, it is the stellar abundances that are the relevant benchmark  \citep{Turrini2021, Pacetti2022}. The assumption of  Solar vs. stellar abundances can significantly affect the inferred planetary abundances, leading to abundance errors larger than the expected JWST atmospheric measurement precision \citep{Greene2016}. The detailed elemental abundances of the host star k2-18 will be beneficial for future JWST analyses to accurately measure the chemical composition of the exoplanet K2-18 b.

The abundance ratios of volatile elements such as C/O play an important role in the location of planet formation within the protoplanetary disk \citep{oberg2011}. A planet with a sub-stellar C/O ratio is likely to have a water-rich atmosphere \citep{Madhusudhan2012, Teske2014} with a formation location within the H$_{2}$O ice line. On the other hand, a planet with a super-stellar C/O ratio is likely to be rich in carbonaceous compounds and have a formation location beyond the H$_{2}$O ice line, which has then experienced an inward  migration  to its current place \citep[e.g.][]{Reggiani2022}. Furthermore, an overabundance of alkali metals, Na and K, has been found in the atmospheres of some hot gas giants relative to their host stars \citep{hands2022}. Such an enhancement of alkali species is thought to be a result of planet formation exterior to the H$_{2}$O ice line followed by inward migration.

However, due to the uncertainties on K2-18~b's internal structure, its C/O ratio has not yet been confidently measured. For example, the observed carbon-bearing species combined with no observed water vapor would imply a relatively high C/O ratio, but this only holds for classical gas-dominated models. If instead, the observed atmosphere is blanketing a planetary ocean, we wouldn't observe any of the water present in the planet and would erroneously infer a high C/O ratio. \citet{Madhusudhan2023} did not present a C/O ratio in their atmospheric observations, and \citet{Wogan2024} assumed a solar C/O ratio in their planetary atmosphere models. As of now, we are unable to measure K2-18~b's C/O ratio with confidence, but hopefully our understanding of the planet and its interior structure will improve with future observations and modeling efforts. This, together with stellar C/O ratio measured in this study, will help us to better understand the formation pathway of the planet.

For our follow-up research, we will attempt to develop an alternative method to determine stellar parameters by performing a deep analysis of parameter sensitivity and the correlation between parameters and elemental abundances. The degeneracy effect is one of the major issues in the spectroscopic determination of stellar parameters, in particular for cool dwarfs. Many spectroscopic studies use inferred values of  one or two parameters from  empirical photometric relations and take them out of synthetic spectral fitting. However, current photometric calibrations may result in unreliable parameter values for some stars, causing large uncertainties in determining the free parameters. One way to overcome this problem is to find the spectral regions/features that are mostly sensitive to only one parameter. Utilizing such collected wavelength intervals will isolate the contribution of each parameter to the respective spectral lines and features during model fitting. This may enable us to determine the input parameters with higher accuracy, which can yield more accurate elemental abundances.

In our future work, We will also apply our abundance measurement technique to other observed cool JWST host stars and measure their chemical abundances, which can then be used to determine the properties of their exoplanet in upcoming JWST analyses.

\

We  wish to thank the anonymous referee for their helpful comments and suggestions, which improved our manuscript. We extend our thanks to Justin Cantrell for his technical support with the high-performance computing system of the physics
and astronomy department, Georgia State University, which was used for this study. N.H. and I.J.M.C. acknowledge support from NSF AAG grant No. 2108686 and from NASA ICAR grant No. NNH19ZDA001N. D.S. thanks the National Council for Scientific and Technological Development – CNPq. T.N. acknowledges support from the Australian Research Council Centre of Excellence for All Sky Astrophysics in 3 Dimensions (ASTRO 3D), through project No. CE170100013. D.S. thanks the National Council for Scientific and Technological DevelopmentCNPq. E.M. acknowledges financial support through a ``Margarita Salas'' postdoctoral fellowship from Universidad Complutense de Madrid (CT18/22), funded by the Spanish Ministerio de Universidades with NextGeneration EU funds.

\clearpage

\appendix
\counterwithin{figure}{section}

\section{Figures}

\begin{figure}[hbt!]
    \centering
    \includegraphics[width=1.0\linewidth]{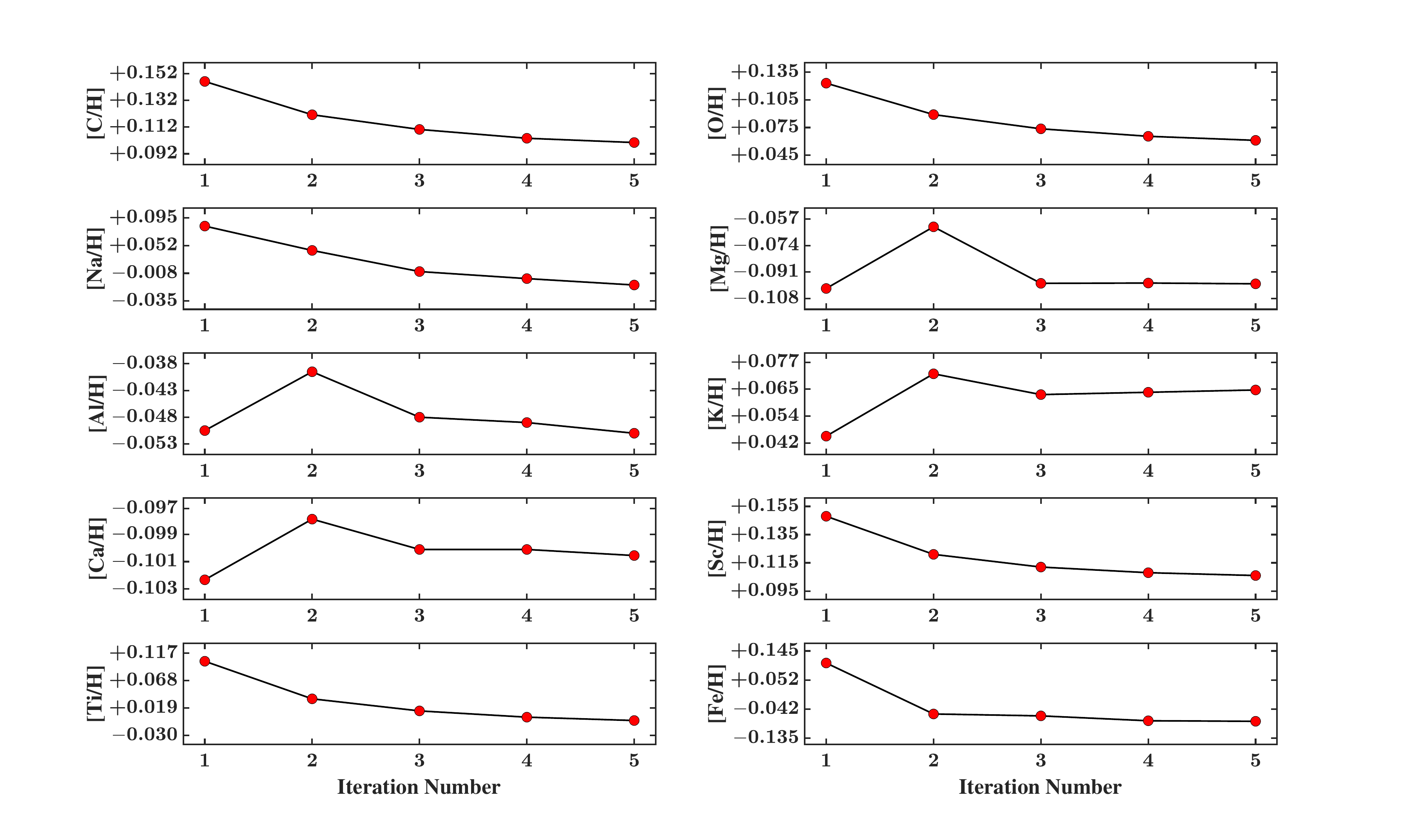}
    \caption{Identical to Figure 3, except that abundances are inferred using the models associated with the deviated effective temperature by +85 K, i.e., $T_{\rm eff}$ = 3632 K, [M/H] = 0.17 dex, log($g$) = 4.90 dex, and $\xi$ = 1.0 km/s. The total number of iterations is 5.}
\label{fig:iteration_temp}
\end{figure}

\begin{figure}[hbt!]
    \centering
    \includegraphics[width=1.0\linewidth]{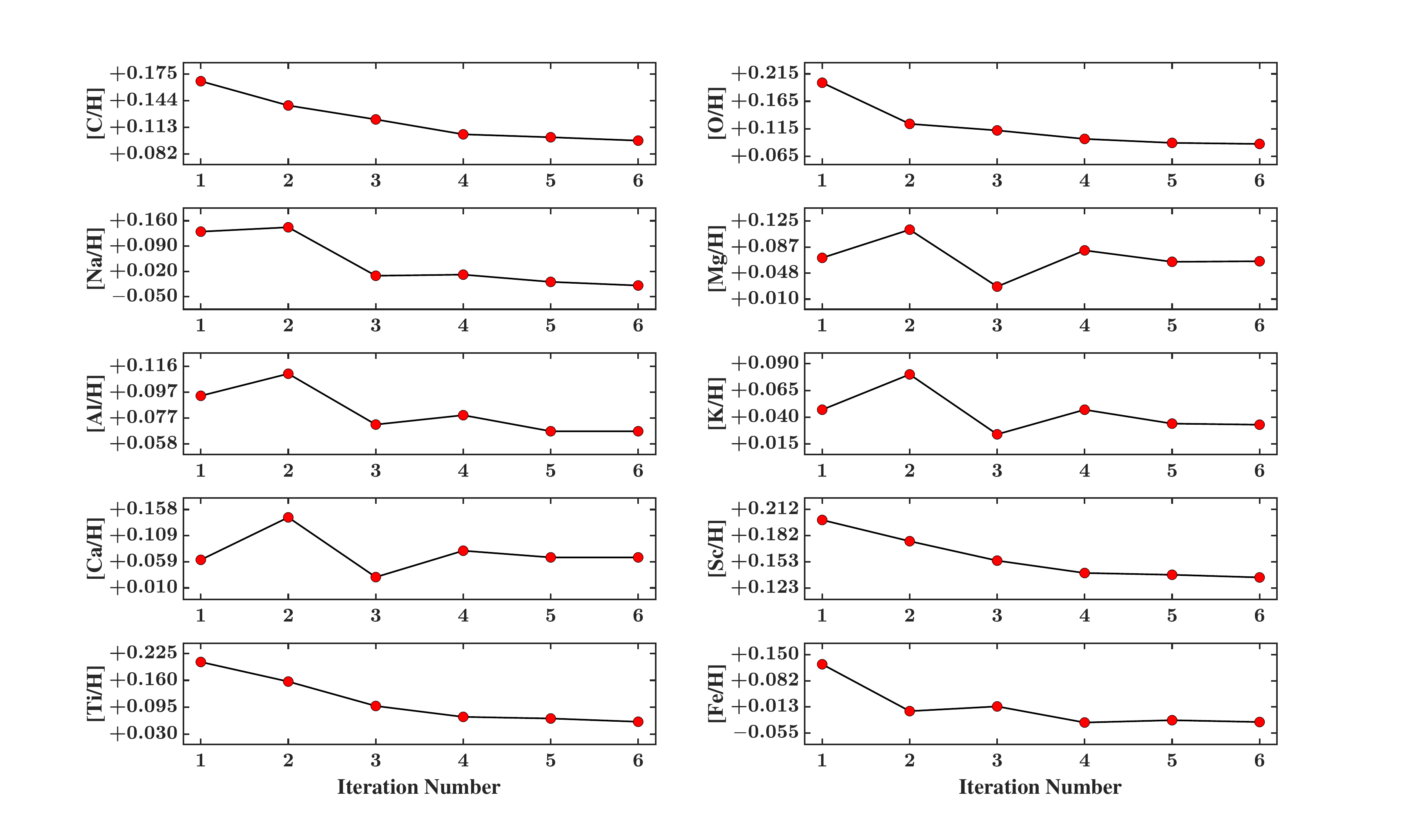}
    \caption{Identical to Figure 3, except that abundances are inferred using the models associated with the deviated overall metallicity by +0.10 dex, i.e., $T_{\rm eff}$ = 3547 K, [M/H] = 0.27 dex, log($g$) = 4.90 dex, and $\xi$ = 1.0 km/s. The total number of iterations is 6.}
\label{fig:iteration_metal}
\end{figure}

\begin{figure}[hbt!]
    \centering
    \includegraphics[width=1.0\linewidth]{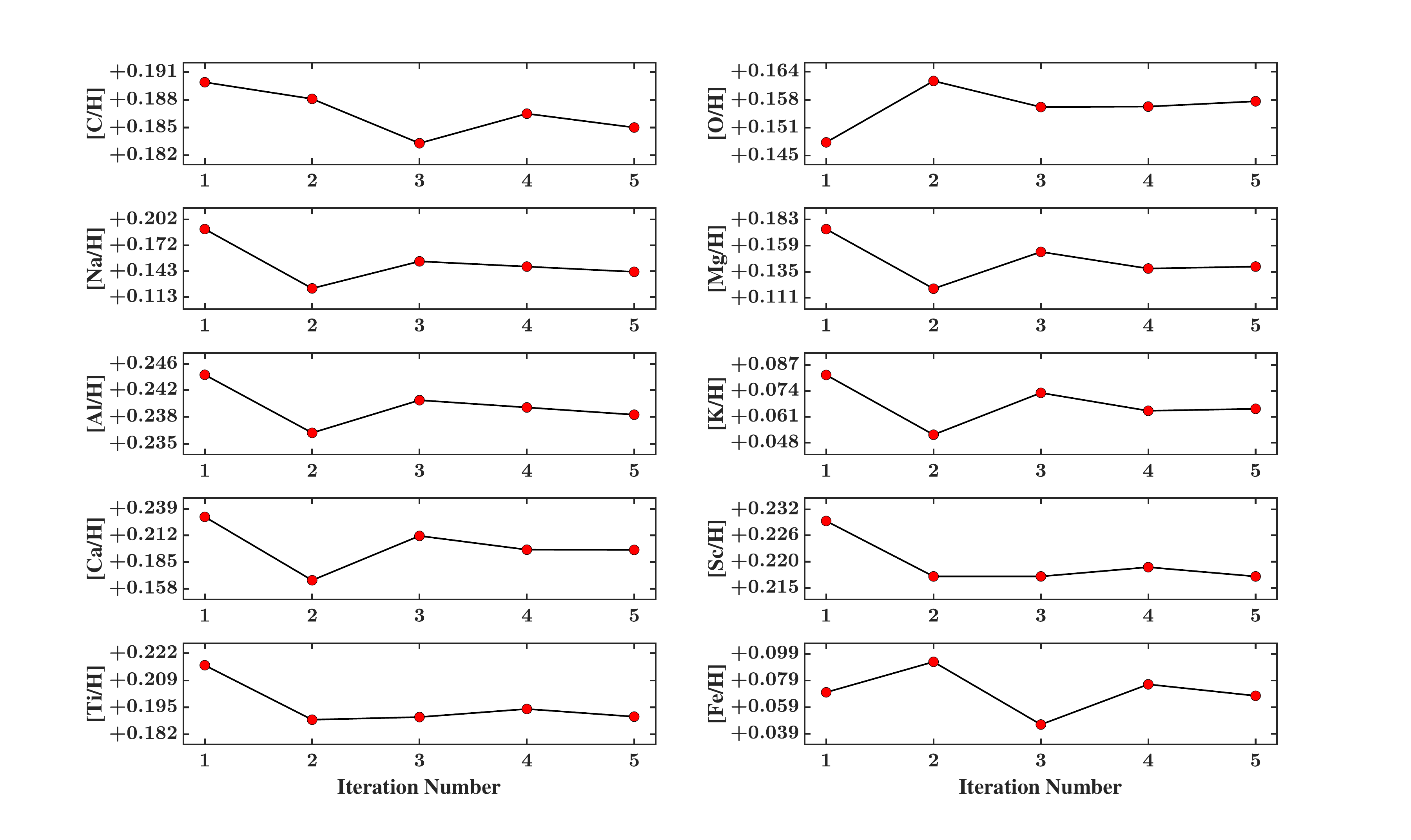}
    \caption{Identical to Figure 3, except that abundances are inferred using the models associated with the deviated surface gravity by +0.10 dex, i.e., $T_{\rm eff}$ = 3547 K, [M/H] = 0.17 dex, log($g$) = 5.00 dex, and $\xi$ = 1.0 km/s. The total number of iterations is 5.}
\label{fig:iteration_grav}
\end{figure}

\begin{figure}[hbt!]
    \centering
    \includegraphics[width=1.0\linewidth]{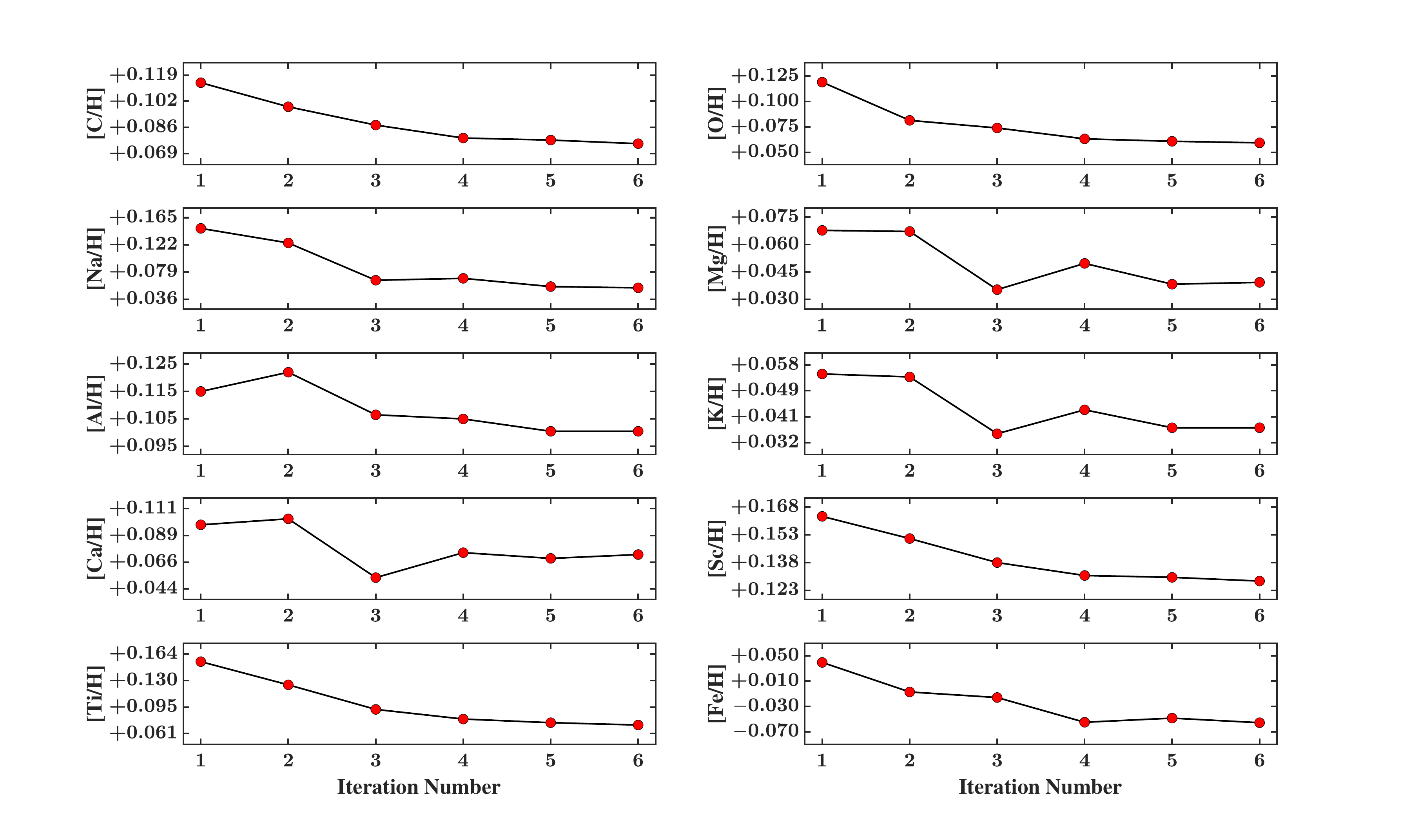}
    \caption{Identical to Figure 3, except that abundances are inferred using the models associated with the deviated microturbulence by +0.10 km/s, i.e., $T_{\rm eff}$ = 3547 K, [M/H] = 0.17 dex, log($g$) = 4.90 dex, and $\xi$ = 1.1 km/s. The total number of iterations is 6.}
\label{fig:iteration_vmic}
\end{figure}

\begin{figure}[hbt!]
    \centering
    \includegraphics[width=1.0\linewidth]{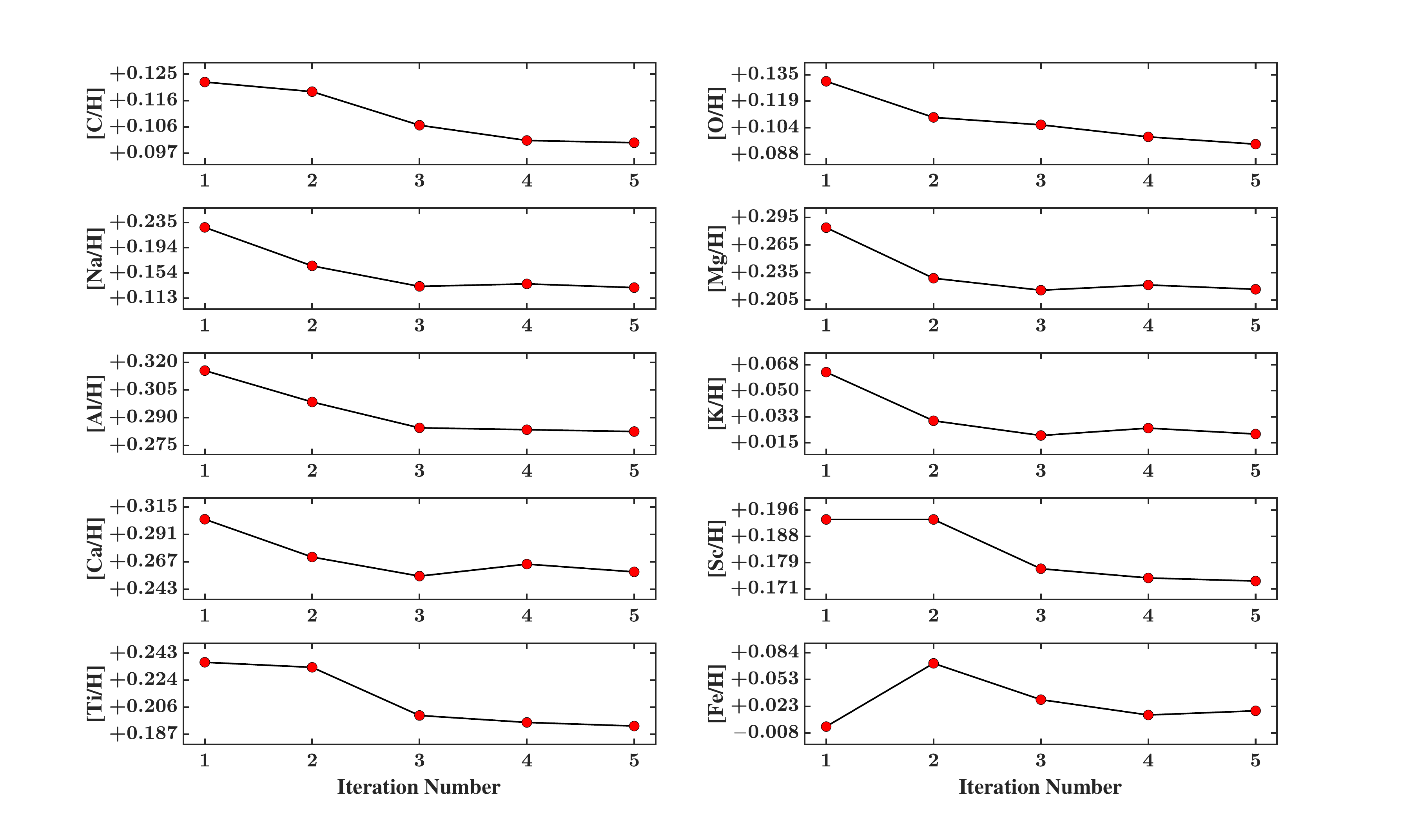}
    \caption{Identical to Figure 3, except that abundances are inferred using the models associated with the deviated effective temperature by $-$85 K, i.e., $T_{\rm eff}$ = 3462 K, [M/H] = 0.17 dex, log($g$) = 4.90 dex, and $\xi$ = 1.0 km/s. The total number of iterations is 5.}
\label{fig:iteration_temp_neg}
\end{figure}

\begin{figure}[hbt!]
    \centering
    \includegraphics[width=1.0\linewidth]{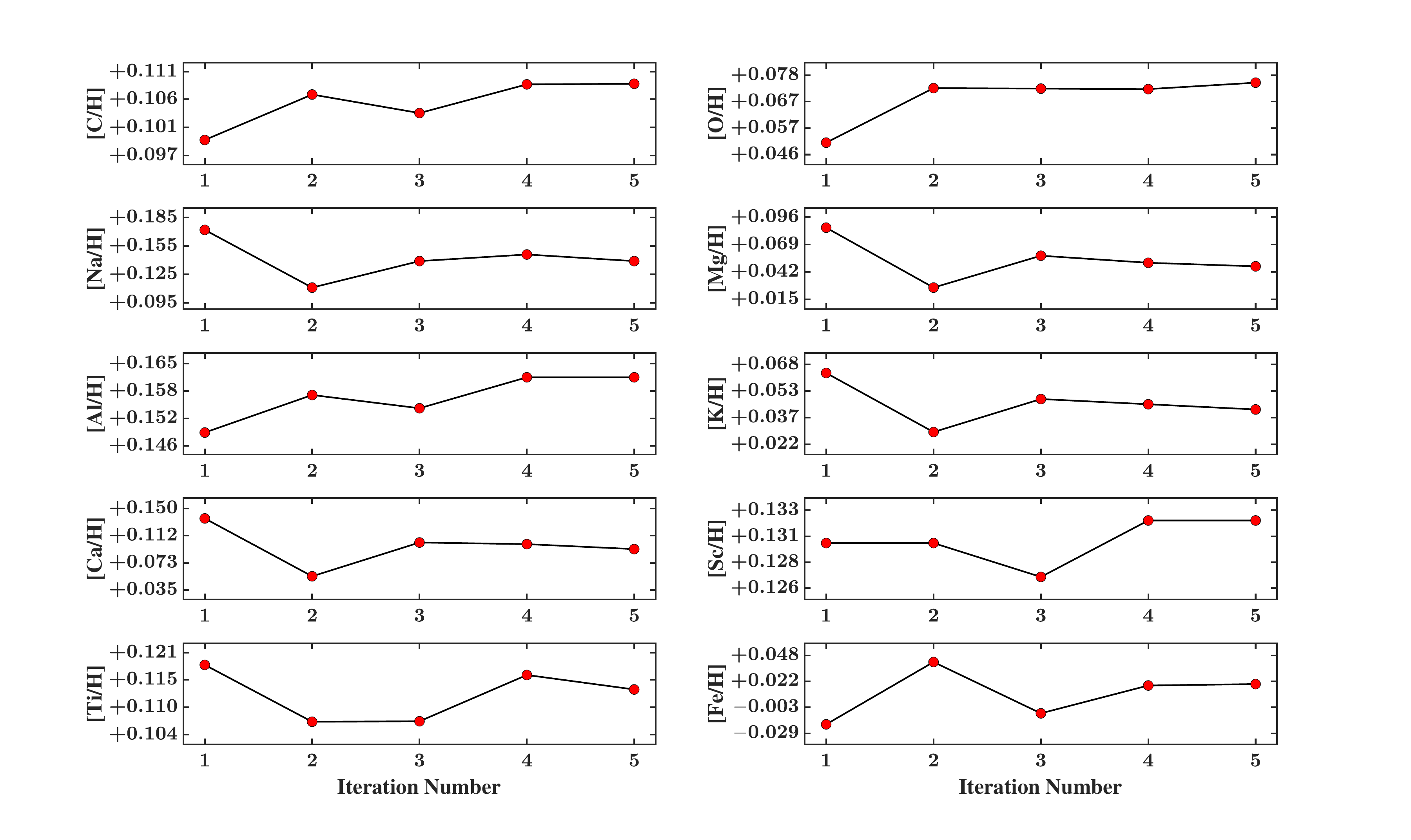}
    \caption{Identical to Figure 3, except that abundances are inferred using the models associated with the deviated overall metallicity by $-$0.10 dex, i.e., $T_{\rm eff}$ = 3547 K, [M/H] = 0.07 dex, log($g$) = 4.90 dex, and $\xi$ = 1.0 km/s. The total number of iterations is 5.}
\label{fig:iteration_metal_neg}
\end{figure}

\begin{figure}[hbt!]
    \centering
    \includegraphics[width=1.0\linewidth]{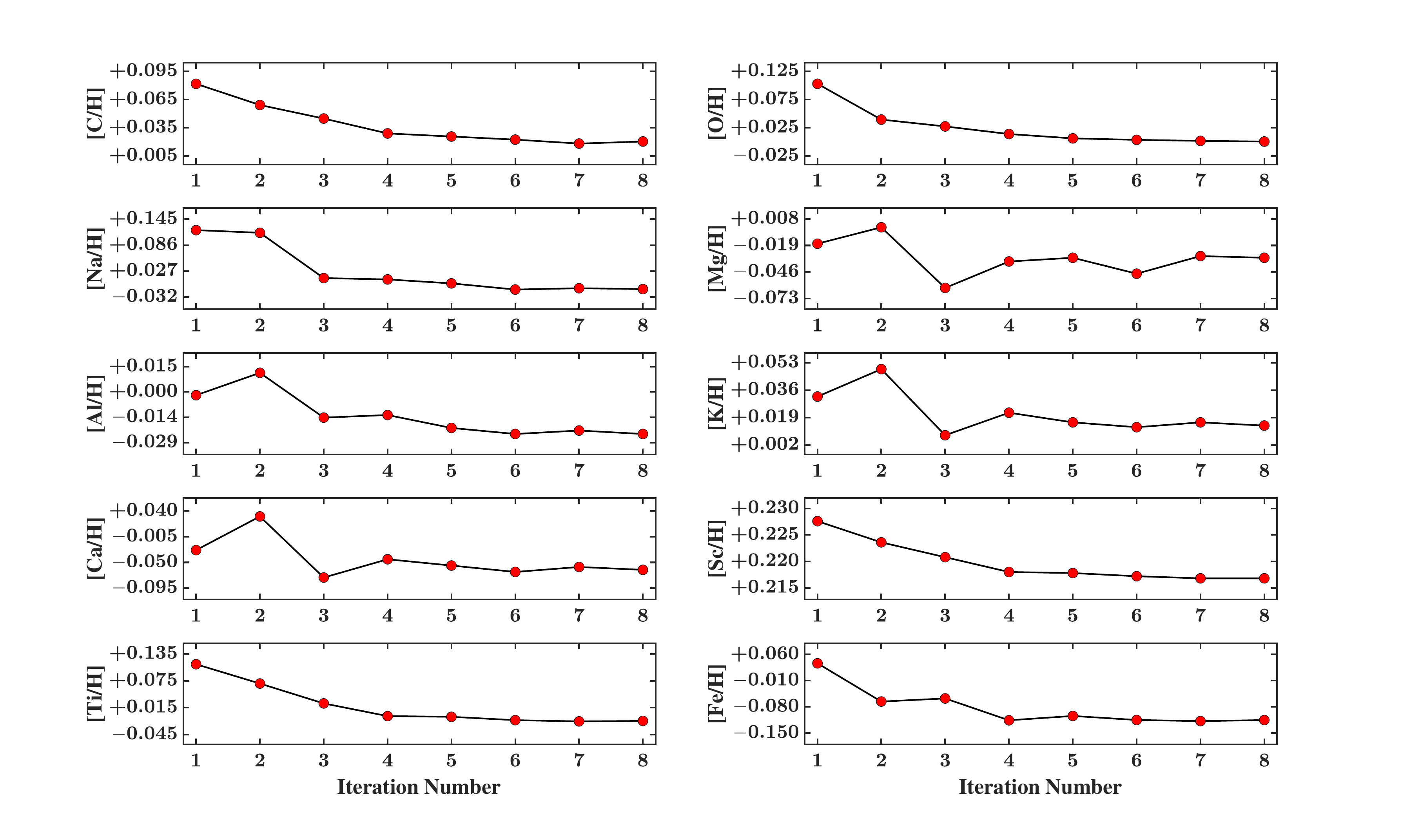}
    \caption{Identical to Figure 3, except that abundances are inferred using the models associated with the deviated surface gravity by $-$0.10 dex, i.e., $T_{\rm eff}$ = 3547 K, [M/H] = 0.17 dex, log($g$) = 4.80 dex, and $\xi$ = 1.0 km/s. The total number of iterations is 8.}
\label{fig:iteration_grav_neg}
\end{figure}

\begin{figure}[hbt!]
    \centering
    \includegraphics[width=1.0\linewidth]{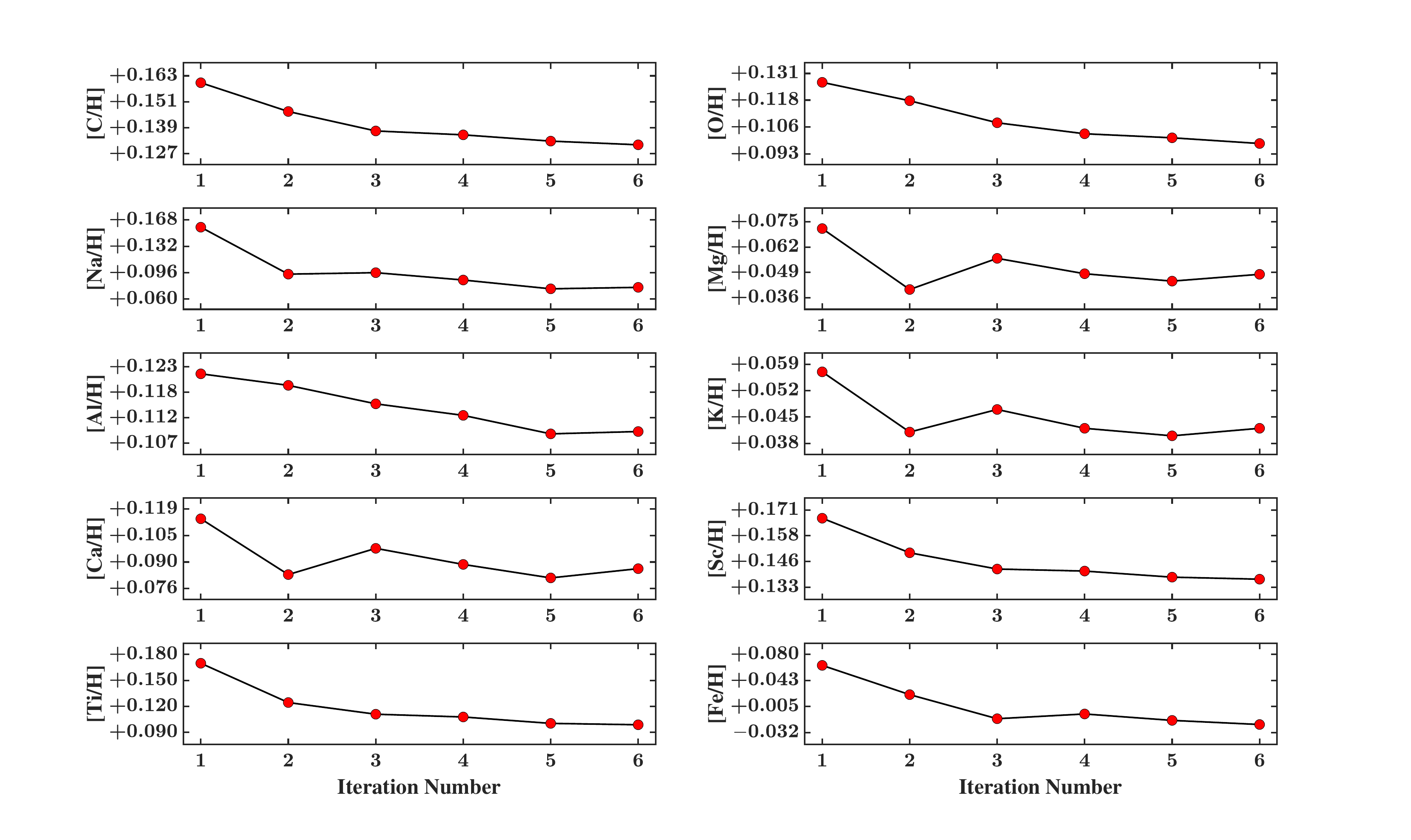}
    \caption{Identical to Figure 3, except that abundances are inferred using the models associated with the deviated microturbulence by $-$0.10 km/s, i.e., $T_{\rm eff}$ = 3547 K, [M/H] = 0.17 dex, log($g$) = 4.90 dex, and $\xi$ = 0.9 km/s. The total number of iterations is 6.}
\label{fig:iteration_vmic_neg}
\end{figure}

{}

\end{document}